\documentclass[twocolumn,amsmath,trackchanges]{aastex702}

\usepackage[dvipsnames]{xcolor}
\usepackage{amssymb}
\usepackage{graphicx}
\usepackage{float}
\let\tablenum\relax
\usepackage{siunitx}
\usepackage{bold-extra}
\usepackage[utf8]{inputenc}

\DeclareRobustCommand{\ion}[2]{\textup{#1\,\textsc{\lowercase{#2}}}}
\DeclareUnicodeCharacter{2060}{\nolinebreak}

\newcommand{\kms}{km\,s$^{-1}$}
\newcommand{\teff}{$T_{\rm eff}$}
\newcommand{\logg}{$\log g$}

\newcommand{\rproc}{$r$-process}
\newcommand{\sproc}{$s$-process}

\newcommand{\AB}[2]{$\mbox{[#1/#2]}$}
\newcommand{\FeH}{\AB{Fe}{H}}
\newcommand{\CFe}{\AB{C}{Fe}}
\newcommand{\BaFe}{\AB{Ba}{Fe}}

\newcommand{\vmic}{$v_\textrm{micr}$}
\newcommand{\AC}{$A(\mathrm{C})$}
\newcommand{\ANa}{$A(\mathrm{Na})$}
\newcommand{\AMg}{$A(\mathrm{Mg})$}

\received{July 14, 2026}
\revised{September 12, 2026}
\accepted{September 18, 2026}
\submitjournal{ApJ}

\begin{document}

\title{The Carbon-Dependent Binary Frequency of CEMP-no Stars} 
\shorttitle{The Carbon-Dependent Binary Frequency of CEMP-no Stars}
\shortauthors{Dixon, J. et al.}

\author[orcid=0000-0001-6168-3130,sname='Dixon']{John D. Dixon}
\affiliation{Department of Physics and Astronomy, Texas A\&M University, College Station, TX 77843, USA}
\affiliation{Mitchell Institute for Fundamental Physics and Astronomy, Texas A\&M University, College Station, TX 77843, USA}
\email{johndixon@tamu.edu}

\author[orcid=0000-0001-6154-8983,sname='Hansen']{Terese T. Hansen}
\affiliation{Department of Astronomy, Stockholm University, AlbaNova University Center, 106 91 Stockholm, Sweden}
\email{terese.hansen@astro.su.se}

\author[orcid=0000-0001-6154-8983,sname='Marshall']{Jennifer Marshall}
\affiliation{Department of Physics and Astronomy, Texas A\&M University, College Station, TX 77843, USA}
\affiliation{Mitchell Institute for Fundamental Physics and Astronomy, Texas A\&M University, College Station, TX 77843, USA}
\email{marshall@tamu.edu}
\author[orcid=0000-0003-4479-1265,sname='Placco']{Vinicius M. Placco}
\affiliation{NSF NOIRLab, Tucson, AZ 85719, USA}
\email{vinicius.placco@noirlab.edu}

\author[orcid=0000-0003-4479-1265,sname='Beers']{Timothy C. Beers}
\affiliation{Department of Physics and Astronomy, University of Notre Dame, Notre Dame, IN 46556, USA}
\affiliation{Joint Institute for Nuclear Astrophysics– Center for the Evolution of the Elements (JINA-CEE), USA}
\email{tbeers@nd.edu}

\author[orcid=0000-0002-0544-2217,sname='Ardern-Arentsen']{Anke Ardern-Arentsen}
\affiliation{Institute of Astronomy, University of Cambridge, Madingley Road, Cambridge CB3 0HA, UK}
\email{anke.arentsen@ast.cam.ac.uk}

\author[orcid=0000-0001-6023-4288,sname='Nordstr\"om']{Birgitta Nordstr\"om}
\affiliation{Stellar Astrophysics Centre, Department of Physics and Astronomy, Aarhus University, Ny Munkegade 120, 8000 Aarhus C, Denmark}
\affiliation{Copenhagen University, The Cosmic Dawn Centre, The Niels Bohr Institute, Jagtvej 128, 2200 Copenhagen, Denmark}
\email{birgitta@nbi.ku.dk}

\author[0000-0003-0183-451X,sname='Aldoroty']{Lauren N. Aldoroty}
\affiliation{University of Maryland Baltimore County, 1000 Hilltop Cir, Baltimore, MD 21250, USA}
\affiliation{NASA GSFC, 8800 Greenbelt Rd, Greenbelt, MD 20771, USA}
\email{laldoroty@umbc.edu}

\author[0000-0002-4657-7679,sname='Cathey']{Jared Cathey}
\affiliation{Department of Astronomy, University of Florida, 211 Bryant Space Sciences Center, Gainesville, FL 32611, USA}
\email{jaredcathey@ufl.edu}

\author[0000-0001-6957-1627,sname='Ferguson']{Peter S. Ferguson}
\affiliation{Physics Department, University of Wisconsin-Madison, 1150 University Avenue, Madison, WI 53706, USA}
\affiliation{DiRAC Institute and the Department of Astronomy, University of Washington, Seattle, WA, USA}
\email{pferguso@uw.edu}

\author[0009-0006-4018-7887,sname='Kristofek']{Katherine Kristofek}
\affiliation{Department of Physics and Astronomy, Texas A\&M University, College Station, TX 77843, USA}
\email{katkristo@tamu.edu}

\author[0000-0002-8892-2573,sname='Lind']{Karin Lind}
\affiliation{Department of Astronomy, Stockholm University, AlbaNova University Centre, Stockholm, Sweden}
\email{karin.lind@astro.su.se}

\author[0009-0004-9457-452X,sname='Malone']{Hudson Malone}
\affiliation{Department of Physics and Astronomy, Texas A\&M University, College Station, TX 77843, USA}
\email{hudsonm33@tamu.edu}

\author[sname='Moreno']{Ferner Moreno}
\affiliation{Department of Physics and Astronomy, Texas A\&M University, College Station, TX 77843, USA}
\email{fern45@tamu.edu}

\author[0000-0002-0004-3057,sname='Myron']{Jessica Myron}
\affiliation{Department of Physics, Montana State University, Bozeman, MT 59717, USA}
\email{jessicamyron1@montana.edu}

\author[0000-0001-5805-5766,sname='Riley']{Alexander H. Riley}
\affiliation{Lund Observatory, Division of Astrophysics, Department of Physics, Lund University, SE-221 00 Lund, Sweden}
\email{alexander.riley@fysik.lu.se}

\author[0000-0003-4501-103X,sname='Starkenburg']{Else Starkenburg}
\affiliation{Kapteyn Astronomical Institute, University of Groningen, Landleven 12, 9747 AD Groningen, The Netherlands}
\email{e.starkenburg@rug.nl}

\author[0000-0002-9762-4308,sname='Webber']{Kaitlin B. Webber}
\affiliation{Department of Physics and Astronomy, Texas A\&M University, College Station, TX 77843, USA}
\affiliation{Mitchell Institute for Fundamental Physics and Astronomy, Texas A\&M University, College Station, TX 77843, USA}
\email{kbwebber@tamu.edu}

\correspondingauthor{John D. Dixon}
\email{johndixon@tamu.edu}

\begin{abstract} 

Studies of the oldest and most metal-poor stars in the Milky Way confirm a common chemical signature of high carbon abundances and subsolar neutron-capture enrichment. The so-called CEMP-no stars are speculated to be bona fide population II stars, potentially tracing the nucleosynthesis of the first stars in the universe. However, constraining the binary nature of CEMP-no stars is crucial for understanding their origins. The binary fraction of CEMP-no stars has tentatively been found to vary with carbon enhancement, \textbf{independent of metallicity}. Here we present the results of radial-velocity monitoring of 30 CEMP-no stars over five years,
\textbf{increasing the total number of CEMP-no stars with a constrained binary status by ${\sim}30\%$ in order to better investigate} the CEMP-no binary fraction as a function of carbon. \textbf{Combining our results with literature data,} we find an overall binary frequency of $50^{+13}_{-13}\%$ among high-carbon ($A(\rm{C})\ge7.3$) CEMP-no stars, compared to $18^{+5}_{-4}\%$ for low-carbon ($A(\rm{C})<7.3$) stars, establishing a statistically significant increase in the CEMP-no binary frequency as a function of carbon for the first time. \textbf{Our results indicate no correlation between metallicity and binary frequency.} Additionally, \textbf{we derive orbital parameters for three new CEMP-no stars in binary systems and estimate secondary masses for these stars as well as seven binaries from the literature. We discuss these mass estimates in the context of investigating} the nature of the unseen companions and the plausibility of a \textbf{mass-transfer formation channel for high-carbon CEMP-no stars}.

\end{abstract}


\section{Introduction} \label{sec:introduction}

The oldest and most metal-poor stars in the Milky Way (MW) provide valuable insights into the chemical history of the early universe \textbf{\citep{beers2005,placco2014,yoon2016,frebel2018,bonifacio2025,yelland2026}}. As metallicity decreases, an increasingly large fraction of MW stars are enhanced in carbon \citep{yong2013,placco2014}; the so-called carbon-enhanced metal-poor (CEMP) stars are defined by \CFe\ $\geq+0.7$ and \FeH\ $\le-1.0$ \citep{beers2005,aoki2007}, but largely fall below \FeH\ $=-2.0$. CEMP stars are further classified based on their abundances of neutron-capture elements, with CEMP-no stars showing no significant enhancement (\BaFe\ $<0$). \textbf{The CEMP-no signature is dominant among the known ultra-metal-poor (\FeH\ $\lesssim-4$) stars in the MW halo \citep{bonifacio2025} and has also been observed in satellite dwarf galaxies \citep{salvadori2015}, making these stars valuable tools to investigate nucleosynthesis and chemical evolution during the infancy of the MW}. This is done by matching abundance patterns of CEMP-no stars to supernova yields. \citet{placco2016b} compared the abundances of 12 stars with \FeH\ $\leq-4$, including 10 CEMP-no stars, to predicted yields from massive metal-free stars \citep{heger2010} and found that the models could only reproduce the abundances in five out of 12 stars, while additional contributions are needed for the remaining seven stars. Other suggested contributors to the CEMP-no signature include faint supernovae \citep{nomoto2006} and stellar winds from rapidly rotating massive stars, so-called ``spinstars" \citep{meynet2006,maeder2015}.

The complex origins of CEMP-no stars become apparent when investigating their absolute carbon abundances (\AC). 
\citet{yoon2016} identified two sub-groups of CEMP-no stars within \AC-\FeH\ phase space: one group (Group~II) in which \AC\ correlates with \FeH\ and another (Group~III) with similar \AC\ values independent of \FeH. These two groups also exhibit different behavior in the \ANa-\AC\ and \AMg-\AC\ spaces, suggesting enrichment from multiple progenitors. More recently, \citet{lee2025} argued for a third subclass (Group~IV) of CEMP-no stars characterized by very high carbon abundances (\AC\ $>7.39$) and extremely low metallicities (\FeH\ $\le-3.1$), for which they tentatively determined a slightly elevated binary frequency of ${\sim}$33\%. These Group~IV stars align with the high-carbon plateau of CEMP stars, which is otherwise largely populated by binary CEMP-$s$ stars formed via mass transfer from an \textbf{asymptotic giant branch (AGB)} companion \citep{lucatello2005,spite2013}; it is not yet known if Group~IV stars have a similar mass-transfer origin.

The binary properties of CEMP-no stars have previously been investigated, with initial results from radial-velocity (RV) monitoring showing that the majority of these stars are single \citep{starkenburg2014,hansen2016a}. Further analysis of larger samples suggested that the binary frequency of CEMP-no stars might vary as a function of \AC, with high-carbon stars more likely to be in binary systems \citep{arentsen2019}. However, previous samples of CEMP-no stars have been too small \textbf{to demonstrate a significant correlation} between binarity and \AC. In this paper, we present the results of RV monitoring of 30 CEMP-no stars. Including literature data, we compile a sample of 90 CEMP-no stars with constrained binary statuses to investigate a potential association between \AC\ and binary frequency. 

The remainder of this paper is structured as follows. Section \ref{sec:data} describes target selection and the collection and reduction of data. Section \ref{sec:derivation_of_rvs} provides details on RV measurements and the determination of binary statuses. Our results are presented in Section \ref{sec:results}, including estimated orbital parameters for three CEMP-no binaries. Section \ref{sec:discussion} discusses the implications of our findings in the context of constraining possible progenitors of the CEMP-no signature. Finally, a summary of our analysis is provided in Section \ref{sec:summary}.


\section{Target selection and observations}\label{sec:data}

\begin{deluxetable*}{lrrrrrrrrrl}
\tablecaption{\label{tab:targets}Details of the CEMP-no Targets Analyzed in this Paper}
\tablehead{\colhead{Star Name} & \colhead{Gaia DR3 ID} & \colhead{$N^a$} & \colhead{RA} & \colhead{DEC} & \colhead{G} & \colhead{\FeH} & \colhead{\AC} & \colhead{\CFe} & \colhead{\BaFe} & \colhead{Ref.} \\
{} & {} & {} & (hh mm ss) & (dd mm ss) & (mag) & {} & {} & {} & {} & {}}
\startdata
HE~0015$+$0048  & 2547143725127991168 & 4 & 00 18 01.43 & $+$01 05 08.1 & 12.94  & $-3.07$ &  $6.65$ &  $+1.29$ & $ -1.17$ & 1, 2  \\
CS~22958$-$042  & 4718427642340545408 & 3 & 02 01 07.43 & $-$57 16 58.8 & 14.40  & $-3.40$ &  $7.59$ &  $+2.56$ & $<-1.02$ & 1, 3   \\
HE~0420$+$0123  & 3279770347306973056 & 1 & 04 23 14.54 & $+$01 30 48.3 & 11.17  & $-3.03$ &  $6.18$ &  $+0.78$ & $ +0.08$ & 1, 2  \\ 
HE~0440$-$1049  & 3184284084625121792 & 1 & 04 42 39.95 & $-$10 43 23.9 & 12.35  & $-3.02$ &  $6.10$ &  $+0.69$ & $ -1.27$ & 1, 4  \\ 
J0918$+$5107    & 1019263054364445184 & 1 & 09 18 52.08 & $+$51 07 21.5 & 12.84  & $-3.11$ &  $6.03$ &  $+0.71$ & $ -0.74$ & 5    \\ 
HE~1012$-$1540  & 3751852536639575808 & 2 & 10 14 53.47 & $-$15 55 53.2 & 13.85  & $-4.17$ &  $6.67$ &  $+2.41$ & $ -0.69$ & 1, 3 \\
J1037$+$2531    &  724816554864554368 & 1 & 10 37 45.90 & $+$25 31 34.2 & 15.03  & $-2.50$ &  $7.03$ &  $+1.10$ &    …     & 6    \\ 
J1054$+$0528    & 3864140775805950208 & 1 & 10 54 33.10 & $+$05 28 12.7 & 12.66  & $-3.30$ &  $5.95$ &  $+0.82$ & $ -0.52$ & 7    \\ 
BS~16077$-$0007 & 4024258569645322112 & 3 & 11 35 17.98 & $+$31 00 23.43& 12.28  & $-2.82$ &  $6.40$ &  $+0.79$ & $ -0.11$ & 1, 8  \\ 
HE~1217$-$0540  & 3584642553499951744 & 1 & 12 19 53.66 & $-$05 57 15.9 & 14.91  & $-2.94$ &  $6.26$ &  $+0.77$ &    …     & 1, 9  \\ 
J1232$-$0545    & 3679882078898432256 & 1 & 12 32 46.06 & $-$05 45 59.1 & 14.03  & $-3.03$ &  $6.18$ &  $+0.78$ & $ -0.51$ & 1, 10  \\ 
HE~1249$-$3121  & 6183795820024064256 & 2 & 12 52 05.14 & $-$31 37 44.9 & 14.17  & $-3.23$ &  $7.02$ &  $+1.82$ &    …     & 1, 9  \\ 
J1253$+$0753    & 3733768078624022016 & 1 & 12 53 46.09 & $+$07 53 43.1 & 12.28  & $-4.02$ &  $6.00$ &  $+1.59$ & $<-0.30$ & 1, 11 \\ 
HE~1310$-$0536  & 3635533208672382592 & 3 & 13 13 31.18 & $-$05 52 12.5 & 14.11  & $-4.15$ &  $6.72$ &  $+2.44$ & $ -0.50$ & 1, 12 \\ 
HE~1311$-$0131  & 3684987592422300800 & 1 & 13 13 42.02 & $-$01 47 15.3 & 12.60  & $-3.15$ &  $6.01$ &  $+0.73$ & $ -0.62$ & 1, 2  \\ 
CS~22877$-$001  & 3621673727165280384 & 1 & 13 13 55.37 & $-$12 11 41.7 & 11.92  & $-3.31$ &  $6.67$ &  $+1.55$ & $ -0.58$ & 1, 3    \\
HE~1330$-$0354  & 3633795224386184448 & 1 & 13 33 10.67 & $-$04 10 05.8 & 14.94  & $-2.29$ &  $7.15$ &  $+1.01$ & $ -0.47$ & 1, 9  \\ 
HE~1338$-$0052  & 3661850088202356992 & 1 & 13 40 59.30 & $-$01 07 20.0 & 15.10  & $-3.06$ &  $6.90$ &  $+1.53$ & $ -0.02$ & 1, 13 \\ 
J1410$-$0555    & 3640973794069231616 & 1 & 14 10 02.83 & $-$05 55 51.8 & 12.35  & $-3.22$ &  $6.77$ &  $+1.53$ & $ -0.10$ & 14   \\ 
HE~1439$-$0218  & 3648798850821202816 & 1 & 14 41 58.30 & $-$02 31 27.0 & 15.83  & $-2.70$ &  $6.60$ &  $+0.87$ & $ -0.15$ & 1, 13 \\ 
J1529$+$0804    & 1164484488577137792 & 3 & 15 29 53.94 & $+$08 04 48.1 & 12.49  & $-3.18$ &  $6.05$ &  $+0.80$ & $ -0.88$ & 6, 14    \\ 
J1556$-$1655    & 6250246214002836608 & 1 & 15 56 28.74 & $-$16 55 33.4 & 13.40  & $-2.79$ &  $6.44$ &  $+0.80$ & $ -0.40$ & 1, 10  \\ 
J1630$+$0953    & 4458577516730343424 & 1 & 16 30 35.82 & $+$09 53 16.9 & 12.97  & $-2.94$ &  $6.75$ &  $+1.26$ & $ -0.79$ & 14, 15   \\ 
J1704$+$1626    & 4546350980218019584 & 1 & 17 04 11.97 & $+$16 26 55.2 & 13.56  & $-2.66$ &  $6.50$ &  $+0.73$ & $ -0.31$ & 5    \\ 
J1709$+$1616    & 4547563844622768768 & 3 & 17 09 59.78 & $+$16 16 13.3 & 14.38  & $-3.71$ &  $6.30$ &  $+1.58$ & $ -0.01$ & 16   \\ 
CS~22950$-$046  & 6876806419780834048 & 1 & 20 21 28.39 & $-$13 16 33.7 & 13.87  & $-4.12$ &  $5.61$ &  $+1.30$ & $ -1.01$ & 1, 3 \\ 
CS~30314$-$067  & 6779790049231492096 & 5 & 20 52 50.99 & $-$34 19 40.5 & 11.46  & $-3.31$ &  $6.70$ &  $+1.58$ & $ -0.55$ & 1, 3  \\
CS~29498$-$043  & 6788448668941293952 & 3 & 21 03 52.12 & $-$29 42 50.3 & 13.35  & $-3.87$ &  $7.62$ &  $+3.06$ & $ -0.51$ & 1, 3  \\
CS~29502$-$092  & 2629500925618285952 & 1 & 22 22 36.00 & $-$01 38 27.6 & 11.61  & $-3.30$ &  $6.59$ &  $+1.46$ & $ -1.46$ & 1, 3  \\
HE~2319$-$5228  & 6501398446721935744 & 3 & 23 21 58.19 & $-$52 11 43.2 & 12.99  & $-2.60$ &  $7.70$ &  $+1.90$ & $ -0.43$ & 1, 17, 18  \\
\enddata
\tablerefs{(1) \citet{yoon2016},           (2) \citet{hollek2011},         (3) \citet{roederer2014},
(4) \citet{hansen2015a},        (5) \citet{bandyopadhyay2024},  (6) \citet{jeong2023},        
(7) \citet{mardini2019a},       (8) \citet{allen2012},          (9) \citet{barklem2005},      
(10) \citet{jacobson2015},      (11) \citet{li2015b},           (12) \citet{hansen2014},      
(13) \citet{cohen2013},         (14) \citet{li2022},            (15) \citet{mardini2019b},    
(16) \citet{li2015a},           (17) \citet{arentsen2019},      (18) \citet{gull2021}}
\tablecomments{Coordinates are obtained from the third data release of Gaia \citep{Gaia_EDR3, Gaia_DR3}. All carbon abundances are corrected for stellar evolution based on \citet{placco2014}.}
\tablenotetext{a}{Number of new radial-velocity observations collected in this work.}
\end{deluxetable*}

Our sample contains 30 CEMP-no stars covering a broad range of \AC\ values, with no particular bias towards high-carbon or low-carbon stars. \textbf{We selected stars that have at least one reported RV measurement from the literature but lack a known or well-constrained binary status.} Twenty-two stars were selected from \citet{yoon2016}, while eight were chosen from various other papers. All targets are presented in Table \ref{tab:targets}, listing the Gaia identifiers, number of observations, coordinates, \textbf{G magnitudes,} \FeH, \AC, \CFe, \BaFe, and literature reference(s) used for target selection.
For this paper, we adopt the CEMP and CEMP-no definitions from \citet{beers2005} and \citet{aoki2007}: the selected targets have abundances \textbf{reported in the literature} that are consistent with \FeH\ $\le-2.0$, \CFe\ $\ge+0.7$, and \BaFe\ $\le0.0$. For all stars, we use the \CFe\ ratio corrected for carbon depletion effects during the later stages of stellar evolution \citep{placco2014}.

All of our targets have \FeH\ $<-2.0$; nine are very metal-poor ($-3.0\le$ \FeH\ $<-2.0$), 17 are extremely metal-poor ($-4.0\le$ \FeH\ $<-3.0$), and four are ultra metal-poor ($-5.0\le$ \FeH\ $<-4.0$) as defined in \citet{beers2005}. Three targets (J1037$+$2531, HE~1217$-$0540, and HE~1249$-$3121) have no published \BaFe\ abundances or upper limits, but we have investigated their Ba abundances using archival spectra (see Appendix Section \ref{sec:appendix_upper_limits}) to verify that they are consistent with \BaFe\ $\le0.0$. 

Our observations consist of 53 high-resolution, low \textbf{signal-to-noise (S/N)} echelle spectra of the target stars, collected with three different instruments. The full list of observations, including previous RVs published in the literature, is given in Appendix Table \ref{tab:appendix_rvs}. In addition, we observe six RV standard stars to \textbf{use as a reference point for RV measurements and} assess the stability of each instrument over the monitoring period, as detailed in Table \ref{tab:standards}.

\subsection{McDonald Observatory Data}
Between September 2020 and May 2025, 24 spectra of 18 stars were obtained with the lower-resolution mode of the Tull Coud\'e Spectrograph \citep[TS23;][]{ts23} on the Harlan J. Smith Telescope at the McDonald Observatory in Fort Davis, TX. All spectra were obtained using the 1\farcs8 slit and 1x1 binning, resulting in a wavelength range of $3400~\rm{\AA}\lesssim\lambda\lesssim10000~\rm{\AA}$, with a spectral resolution of $R\sim44,000$ at 5000~\AA.
TS23 spectra were reduced using standard data reduction routines from the \texttt{echelle} package of the open-source image reduction software \texttt{IRAF~v2.18}\footnote{\href{https://iraf.noirlab.edu/}{https://iraf.noirlab.edu/}} \citep{iraf1,iraf2,iraf3}, released and managed by NOIRLab. This process included bias subtraction, cosmic ray removal, flat-field correction, scattered-light subtraction, optimal aperture extraction, and application of a wavelength solution using a ThAr calibration lamp. At 5100~\AA, the S/N \textbf{per pixel} of these spectra ranges from ${\sim}6$--15.

\subsection{Las Cumbres Observatory Data}
In February 2022, 3 stars (HE~0420$+$0123, HE~0440$-$1049, and J1410$-$0555) were each observed once with the Network of Robotic Echelle Spectrographs \citep[NRES;][]{NRES} on 1.0-meter telescopes located at the Wise Observatory and the Cerro Tololo Interamerican Observatory, operated by the Las Cumbres Observatory Global Telescope network \citep[LCOGT;][]{LCOGT}. With a fiber width of 2\farcs77 and 1x1 binning, this setup provides a wavelength range of $3900~\rm{\AA}\lesssim\lambda\lesssim9000~\rm{\AA}$, with a spectral resolution of $R\sim42,000$ at 5000~\AA.
All NRES spectra were reduced with \texttt{IRAF} as described above for the TS23 spectra. The S/N \textbf{per pixel} at 5100~\AA\ for these spectra ranges from ${\sim}5$--10.

\subsection{La Silla Observatory Data}
Between October 2019 and March 2020, 26 spectra of 10 stars were obtained with the Fiber-fed Extended Range Optical Spectrograph \citep[FEROS;][]{feros,feros_data} mounted on the MPG/ESO 2.2-meter telescope at La Silla Observatory, as part of ESO program 0104.A-9008(A). With a fiber diameter of 2\farcs0 and 1x1 binning, this setup provides a wavelength range of $3500~\rm{\AA}\lesssim\lambda\lesssim9200~\rm{\AA}$, with a spectral resolution of $R\sim45,000$ at 5000~\AA.
All FEROS spectra were reduced using the FEROS Data Reduction System (DRS) within the publicly available ESO-MIDAS image processing system. The S/N \textbf{per pixel} at 5100~\AA\ is ${\sim}$15--20 for these spectra.

\subsection{Literature Data}

We supplement the data described above with observations from the literature, largely collected using the SIMBAD \citep{simbad} and SAGA \citep{saga} databases. In particular, we compile an additional 83 RVs for our 30 target stars. In addition to our targets, radial-velocity data exist for 62 CEMP-no stars in the literature, sufficient to determine their binary nature. These 62 stars are presented in Appendix Table \ref{tab:literature_stars}, along with their Gaia DR3 identifiers, coordinates, \FeH, \AC, \BaFe, binary statuses, and references.

\begin{deluxetable*}{lcrrrrrrrrrc}[ht!]
\tablecaption{\label{tab:standards} \textbf{Radial-Velocity Standard Stars Observed in this Program}}
\tablehead{\colhead{Star Name} & Instrument & \colhead{$N^a$} &  \colhead{RA} & \colhead{DEC} & \colhead{G} & \FeH & \colhead{$\Delta T^b$} & \colhead{$v_r$} & $\sigma_{v_r}$ & \colhead{Gaia RV} & \colhead{Ref.} \\
{} & {} & {} & (hh mm ss) & (dd mm ss) & (mag) & {} & {} & (\kms) & (\kms) & (\kms) & {}
}
\startdata
HD~38230  &  TS23  & 6  & 05 46 01.89  & $+$37 17 04.7  & 7.12 & $-0.11$ &   1118   & $-28.86$  & 0.46     & $-29.21\pm0.13$   & 1   \\
HD~80536  &  TS23  & 5  & 09 21 10.60  & $+$25 09 47.6  & 7.12 &  $0.05$ &     42   & $-38.13$  & 0.21     & $-38.02\pm0.14$   & 1   \\
HD~122563 &  TS23  & 7  & 14 02 31.85  & $+$09 41 09.9  & 5.87 & $-2.79$ &   1856   & $-26.36$  & 0.57     & $-26.13\pm0.12$   & 2   \\
HD~182488 &  TS23  & 8  & 19 23 34.01  & $+$33 13 19.1  & 6.16 &  $0.14$ &   1710   & $-21.47$  & 0.30     & $-21.62\pm0.12$   & 1   \\
\hline
HD~38230  &  NRES  & 1  & 05 46 01.89  & $+$37 17 04.7  & 7.12 & $-0.11$ &      0   & $-29.42$  & $0.50^c$ & $-29.21\pm0.13$   & 1   \\
HD~122563 &  NRES  & 1  & 14 02 31.85  & $+$09 41 09.9  & 5.87 & $-2.79$ &      0   & $-26.02$  & $0.50^c$ & $-26.13\pm0.12$   & 2   \\
HD~182488 &  NRES  & 1  & 19 23 34.01  & $+$33 13 19.1  & 6.16 &  $0.14$ &      0   & $-21.34$  & $0.50^c$ & $-21.62\pm0.12$   & 1   \\
\hline
HD~97343  & FEROS  & 3  & 11 12 01.19  & $-$26 08 12.0  & 6.86 & $-0.01$ &     64   & $39.78$   & 0.07     &  $39.76\pm0.12$   & 3   \\
HD~217357 & FEROS  & 3  & 23 00 16.12  & $-$22 31 27.6  & 7.26 &  $0.14$ &     53   & $16.10$   & 0.29     &  $15.85\pm0.14$   & 4
\enddata
\tablerefs{Gaia RVs obtained from \citet{Gaia_DR3_RVs}. Metallicities obtained from (1) \citet{boeche2016}, (2) \citet{lombardo2022}, (3) \citet{luck2018}, and (4) \citet{soubiran2022}}
\tablenotetext{a}{Number of observations taken with the specified instrument.}
\tablenotetext{b}{Monitoring period of the star (in days) with the specified instrument.}
\tablenotetext{c}{\textbf{A conservative estimate of 0.50~\kms\ is adopted as the instrumental uncertainty of NRES, as our data are insufficient to calculate RV variance.}}
\end{deluxetable*}

\section{Measurement of Radial Velocities}\label{sec:derivation_of_rvs}

RVs for all program and standard stars were determined with either the \textbf{\texttt{rvidlines} or \texttt{fxcor} package from \texttt{IRAF}, as detailed below. The \texttt{rvidlines} package measures an RV from a spectrum based on the positions of strong spectral features.} The \texttt{fxcor} package applies a Fourier cross-correlation of a stellar spectrum with a template spectrum, and determines a Doppler shift by fitting a Gaussian curve to the cross-correlation peak. \textbf{All reported RVs in this work are heliocentric RVs.}

\subsection{Program Stars}\label{sec:program_stars}
{\bf Heliocentric RVs of the target stars are determined via order-by-order cross-correlation against a high S/N spectrum of an RV standard star. Orders dominated by telluric features were discarded automatically, while the full wavelength range was used for the remaining echelle orders.} 
For consistency, all RV measurements of a program star taken with the same instrument use the same template spectrum (listed in Table \ref{tab:chi2}). 
Depending on the metallicity of the program star and the S/N of the spectrum, between 3 and 40 echelle orders were usable for cross-correlation. The three orders covering the \ion{Mg}{I}~b triplet (5167, 5173, and 5184~\AA), infrared \ion{Ca}{II} triplet (8498, 8542, and 8662~\AA), and H$\beta$ line (4861~\AA) were used for all stars. 
\textbf{We then calculated a weighted-mean RV, using the Tonry-Davis ratio \citep{tonry1979} reported by \texttt{IRAF} to weight each order by reliability, and employing a $\sigma$-clipping algorithm to discard orders that fall more than $\pm3\sigma$ from the mean.}
\textbf{By examination of the order-by-order RVs of our spectra, we do not find any trends with wavelength that could suggest errors in the wavelength solution.}

\subsection{Radial-Velocity Standard Stars}\label{sec:standard_stars}
\textbf{The RV standard stars used for this work are listed in Table \ref{tab:standards}, and were selected from radial velocity catalogs \citep{udry1999,soubiran2013,Gaia_DR3_RVs}.
While these standards differ somewhat in chemical composition from CEMP-no stars, they are similar in spectral type to our targets.}
\textbf{For each instrument, we used \texttt{rvidlines} to determine the RV of the highest S/N spectrum. We then measured RVs of all other standard star spectra from the same instrument with \texttt{fxcor}, using the highest S/N spectrum as the template.}

We observed four RV standard stars with TS23 between April 2020 and May 2025 to monitor the stability of the instrument. Using our RV measurements of these stars, we calculate an overall RV variance of 0.42~\kms, weighting the variance of each standard star by the number of observations. We adopt this as the systematic RV uncertainty due to instrumental shifts in TS23.
Two RV standard stars were observed with FEROS between October 2019 and March 2020. Using our RV measurements of these stars, we calculate an overall variance of 0.21~\kms, which we adopt as the instrumental uncertainty for FEROS during the monitoring period. 
\textbf{For NRES, three RV standards were observed only once each between February 2022 and April 2022. We cannot estimate instrumental variance from just one RV, but since our heliocentric RVs for all three stars are well within 0.50~\kms\ of the RVs reported by Gaia, we adopt a conservative estimate of 0.50~\kms\ for the instrumental uncertainty of NRES.}

\textbf{Table \ref{tab:standards} also lists the Gaia DR3 RVs and G magnitudes for each standard star. All standard stars are bright ($G<7.3$) with a re-normalized unit weight error (RUWE) less than 1.10 \citep{GaiaDR3_astrometry}, and our measured RVs are consistent with the Gaia values. Thus, we adopt these Gaia RVs for the purposes of cross-correlation with target spectra, and we add our estimated instrumental uncertainties in quadrature with the measurement uncertainty on each RV to achieve the values reported in this work.}

\subsection{Determination of Binarity}\label{sec:identifying_binaries}

We assess RV variability (and thereby the binary nature) of each CEMP-no star using a standard $\chi^2$ statistical test:

\begin{displaymath}
    \chi^2 = \sum_{i=1}^{n}\left(\frac{v_{i}-v_{0}}{\sigma_{i}}\right)^2
\end{displaymath}

\noindent where $v_i$ is an RV measurement, $\sigma_i$ is its uncertainty, and $v_0$ is the weighted mean of all RVs. 
We consider a $p$-value of $p(\chi^2)<0.01$ as evidence of RV variation. \textbf{Since our analysis includes previous RVs from the literature, this criterion was chosen to be slightly stricter than a $2\sigma$ confidence level to account for the potential inhomogeneity of RV measurements. All target stars have a distance greater than 250~pc, and a majority are farther than 1~kpc \citep{GaiaDR3_astrometry}. Thus, the Gaia RUWE is not a reliable indicator of binarity \citep{castro-ginard2024}.} When available, we provide the most recently published RV from Gaia \citep{Gaia_DR2,Gaia_DR2_RVs,Gaia_DR3,Gaia_DR3_RVs}, as a reference point in Appendix Table \ref{tab:appendix_rvs}. \textbf{However, we exclude these values from $\chi^2$ analysis as they are derived from an unknown number of observations.}

\section{Results}\label{sec:results}

\begin{deluxetable*}{llrrrrrrc}
\tablecaption{\label{tab:chi2} \textbf{Mean Heliocentric RVs and $\chi^2$ Analysis for All Program Stars}}
\tablehead{\colhead{Star ID} & \colhead{Template} & \colhead{$N^a$} & \colhead{$\Delta T^b$} & \colhead{\AC} & \colhead{Weighted RV} & \colhead{$\chi^2$} & \colhead{$p(\chi^2)$} & \colhead{Binary?} \\
{} & {} & {} & {} & {} & (\kms) & {} & {} & {}
} 
\startdata
HE~0015$+$0048  &      HD~97343  &  6 &  $  4488$  & $6.65$ &  $ -37.53\pm1.41$  &     18.1301 &    0.00279 &  {Yes${}^c$} \\
CS~22958$-$042  &      HD~97343  &  8 &    $6645$  & $7.59$ &  $ 167.14\pm4.36$  &     96.5065 & $<10^{-5}$ &  {Yes} \\
HE~0420$+$0123  &     HD~122563  &  7 &    $5611$  & $6.18$ &  $ -55.80\pm1.00$  &      5.6885 &    0.45898 &  {No}  \\
HE~0440$-$1049  &      HD~38230  &  2 &    $5961$  & $6.10$ &  $ 137.58\pm0.23$  &      0.0189 &    0.89072 &  {No}  \\
J0918$+$5107    &     HD~182488  &  3 &  $  5619$  & $6.03$ &  $ -51.04\pm0.98$  &      0.9722 &    0.61503 &  {No}  \\
HE~1012$-$1540  &      HD~97343  & 14 &    $6530$  & $6.67$ &  $ 225.56\pm0.48$  &     12.0672 &    0.52214 &  {No${}^c$} \\
J1037$+$2531    &     HD~182488  &  2 &  $  1952$  & $7.03$ &  $  28.74\pm0.07$  &      0.0004 &    0.98323 &  {No}  \\
J1054$+$0528    &     HD~182488  &  2 &  $  3650$  & $5.95$ &  $  82.31\pm0.16$  &      0.0874 &    0.76756 &  {No}  \\
BS~16077$-$0007 &      HD~38230  &  5 &  $  7302$  & $6.40$ &  $ -37.93\pm1.14$  &     29.8727 &    0.00001 &  {Yes}  \\
HE~1217$-$0540  &      HD~80536  &  2 &  $  7977$  & $6.26$ &  $ 149.62\pm1.58$  &      0.4179 &    0.51797 &  {No}  \\
J1232$-$0545    &     HD~182488  &  2 &    $3931$  & $6.18$ &  $ 291.47\pm1.85$  &      0.6217 &    0.43042 &  {}  \\
HE~1249$-$3121  &      HD~97343  &  3 &  $  6159$  & $7.02$ &  $ 223.20\pm1.31$  &      2.3846 &    0.30352 &  {No}  \\
J1253$+$0753    &     HD~122563  &  2 &  $  3996$  & $6.00$ &  $  72.57\pm1.45$  &      1.3289 &    0.24901 &  {No}  \\
HE~1310$-$0536  & Multiple${}^d$ &  6 &    $6939$  & $6.72$ &  $ 110.14\pm0.87$  &      4.4106 &    0.49193 &  {No}  \\
HE~1311$-$0131  &     HD~182488  &  3 &    $6219$  & $6.01$ &  $ 125.17\pm0.32$  &      0.0692 &    0.96599 &  {No}  \\
CS~22877$-$001  &      HD~97343  &  5 &    $6209$  & $6.67$ &  $ 167.36\pm1.37$  &     12.0672 &    0.01926 &  {No${}^c$} \\
HE~1330$-$0354  &      HD~80536  &  2 &  $  7984$  & $7.15$ &  $  35.63\pm0.16$  &      0.0132 &    0.90857 &  {No}  \\
HE~1338$-$0052  &      HD~80536  &  2 &    $7694$  & $6.90$ &  $ -39.13\pm0.10$  &      0.0024 &    0.96081 &  {No}  \\
J1410$-$0555    &      HD~38230  &  2 &  $  2849$  & $6.77$ &  $  93.09\pm1.42$  &      1.7856 &    0.18147 &  {No}  \\
HE~1439$-$0218  &      HD~80536  &  2 &    $7694$  & $6.60$ &  $ 213.74\pm0.47$  &      0.0445 &    0.83287 &  {No}  \\
J1529$+$0804    &     HD~122563  &  5 &  $  3696$  & $6.05$ &  $  23.86\pm1.27$  &     13.4221 &    0.00939 &  {Yes}  \\ 
J1556$-$1655    &     HD~122563  &  3 &    $4374$  & $6.44$ &  $ 34.34\pm27.98$  &   1913.6935 & $<10^{-5}$ &  {Yes}  \\
J1630$+$0953    &     HD~122563  &  3 &  $  3735$  & $6.75$ &  $  58.00\pm0.40$  &      0.5732 &    0.75083 &  {No}   \\
J1704$+$1626    &     HD~182488  &  2 &  $  1883$  & $6.50$ &  $-176.36\pm2.99$  &      3.6919 &    0.05468 &  {}     \\
J1709$+$1616    &     HD~182488  &  4 &  $  4273$  & $6.30$ &  $-340.81\pm2.05$  &      3.6474 &    0.30214 &  {No}   \\
CS~22950$-$046  &     HD~182488  &  4 &    $9031$  & $5.61$ &  $ 108.68\pm0.59$  &      0.4623 &    0.79364 &  {No}   \\
CS~30314$-$067  &      HD~97343  &  6 &  $  2522$  & $6.70$ &  $ 146.22\pm0.49$  &      1.8519 &    0.86924 &  {No${}^c$} \\
CS~29498$-$043  &      HD~97343  & 13 &  $  6996$  & $7.62$ &  $ -32.49\pm0.35$  &      6.8116 &    0.86981 &  {No${}^c$} \\
CS~29502$-$092  &      HD~97343  & 13 &  $  6974$  & $6.59$ &  $ -67.13\pm0.56$  &     18.2030 &    0.10966 &  {No${}^c$} \\
HE~2319$-$5228  &      HD~97343  &  3 &  $  1979$  & $7.70$ &  $ 284.62\pm5.66$  &     78.2775 & $<10^{-5}$ &  {Yes${}^c$} \\
\enddata
\tablecomments{Listed RV uncertainties are the variance-weighted standard deviation of all measurements used for statistical analysis, while individual RV uncertainties are provided in Appendix Table \ref{tab:appendix_rvs}.}
\tablenotetext{a}{Total number of RVs used to determine a binary status.}
\tablenotetext{b}{\textbf{Monitoring period based on all available observation dates.}}
\tablenotetext{c}{Binary status determined from previous literature and confirmed by our RV measurements.}
\tablenotetext{d}{This star has observations from both TS23 and FEROS; TS23 observations use the HD~182488 template and FEROS observations use the HD~97343 template.}
\end{deluxetable*}

Variance-weighted mean heliocentric RVs for the target stars are listed in Table \ref{tab:chi2}, along with the monitoring period, number of observations, template spectrum, \AC, results of $\chi^2$ analysis, and our determined binary status for each star. For seven stars with binary statuses already provided in the literature, our analysis confirms the pre-existing labels.

In total, we present 53 RV measurements of 30 CEMP-no stars obtained over five years. Detailed results of our individual RV measurements are presented in Appendix Table \ref{tab:appendix_rvs}, including the Julian date (HJD) of the observation, S/N \textbf{per pixel} at $\lambda=5100~\rm{\AA}$, number of echelle orders used for the RV measurement, and the weighted-mean RV from order-by-order Fourier cross-correlation of each observation.
The RV uncertainties range from approximately 0.5~\kms\ (for brighter targets) to 3.3~\kms\ (for very faint, metal-poor, or hot stars with fewer detectable spectral lines).

\subsection{New CEMP-no Binary Statuses}\label{sec:binaries}

In total, we identify four new binary systems and confirm the status of two binaries identified by previous literature. Twenty-two stars exhibit no evidence of RV variation, and the status of two stars remains uncertain, as detailed in Table \ref{tab:chi2}.
Since we use literature data as well as observations from different telescopes, our data are not homogeneous, and instrumental differences may introduce non-physical variation into our results. Thus, the $\chi^2$ test may overestimate the likelihood of variability, as mentioned in \citet{hansen2016a}, so we interpret our results with caution when \textbf{assessing the binary status of stars with $p(\chi^2)$ values slightly below 0.01}. 

For J1556$-$1655, our measured RV of $40.28\pm0.55$~\kms\ is close to the published value of $41.374\pm1.162$~\kms\ from \citet{buder2021} and the Gaia RV of $38.65\pm2.89$~\kms\ \citep{Gaia_DR3_RVs}, but differs by ${\sim}90$~\kms\ from the $-50.1\pm2.0$~\kms\ reported by \citet{jacobson2015}. This could possibly indicate a long-period, highly eccentric binary, but more RV data are needed to constrain an orbit.
For BS~16077$-$0007, $p(\chi^2)\simeq10^{-5}$ due to low RV uncertainties, despite all reported RVs falling within a 5~\kms\ range. J1529$+$0408 has a higher $p(\chi^2)$ value of 0.00784, \textbf{indicating variations at the $2\sigma$ level}; all RV measurements for this star fall within 4~\kms\ of each other. However, we fit orbits to RV measurements of BS~16077$-$0007 and J1529$+$0408 (see Section \ref{sec:orbits}), suggesting they belong to binary systems.

\textbf{One star in our sample (CS~22958$-$042) was speculated by \citet{sivarani2006} to be an extremely short-period binary, as two RV measurements from the same night differed by ${\sim}$3~\kms, but further observations did not find similar RV variations \citep{roederer2014}. Our three measurements of this star also vary by 3~\kms, but due to larger uncertainties, we cannot confirm any short-term variability. However, our RVs differ by ${\sim}10$~\kms\ from the RVs measured two decades ago, signaling long-term RV variation in this star. Thus, we declare this star as binary due to evidence for the presence of at least one binary companion.}

Some stars labeled as single may actually be undetected long-period, highly eccentric, or face-on binary systems. This is especially true for eight single stars in our sample with only two RV measurements, including this work (HE~0440$-$1049, J1037$+$2531, HE~1217$-$0540, J1253$+$0753, HE~1330$-$0354, HE~1338$-$0052, J1410$-$0555, and HE~1439$-$0218). \textbf{Additionally, for some stars, certain epochs may be oversampled due to clustered observation dates, affecting the reliability of the $\chi^2$ analysis. Three binaries (HE~0015$+$0048, CS~22958$-$042, and HE~2319$-$5228) and six single stars (HE~0420$+$0123, HE~1012$-$1540, CS~22877$-$001, CS~30314$-$067, CS~29498$-$043, and CS~29502$-$092) have multiple RVs measured within a few days of each other, but for each of these stars, our determined binary status remains unchanged if each cluster is treated as a single weighted-mean RV.}

Finally, for J1232$-$0545 and J1704$+$1626, the binary status (single) is somewhat uncertain. \citet{dietz2020} reported an RV of $284.875\pm8.530$ for J1232$-$0545, compatible with our measurement of $291.732\pm1.672$~\kms\ and the Gaia RV of $293.96\pm4.64$~\kms; however, the larger Gaia uncertainty may indicate RV variation. For J1704$+$1626, we measure an RV of $-177.20\pm1.15$~\kms, while \citet{bandyopadhyay2024} finds $-171.0$~\kms, leading to a marginal $p(\chi^2)$ value of 0.055. However, the RV reported in \citet{bandyopadhyay2024} has no listed uncertainty. For the purposes of the $\chi^2$ analysis, we assume a conservative uncertainty of 3~\kms\ when no literature value is provided, but for this particular star, $p(\chi^2)$ is highly dependent on the unknown RV uncertainty, so it is difficult to declare a binary status. We leave the statuses of these two stars as unknown and note that future RV monitoring is warranted.

\begin{figure*}[t!]
\centering
\includegraphics[scale=0.73]{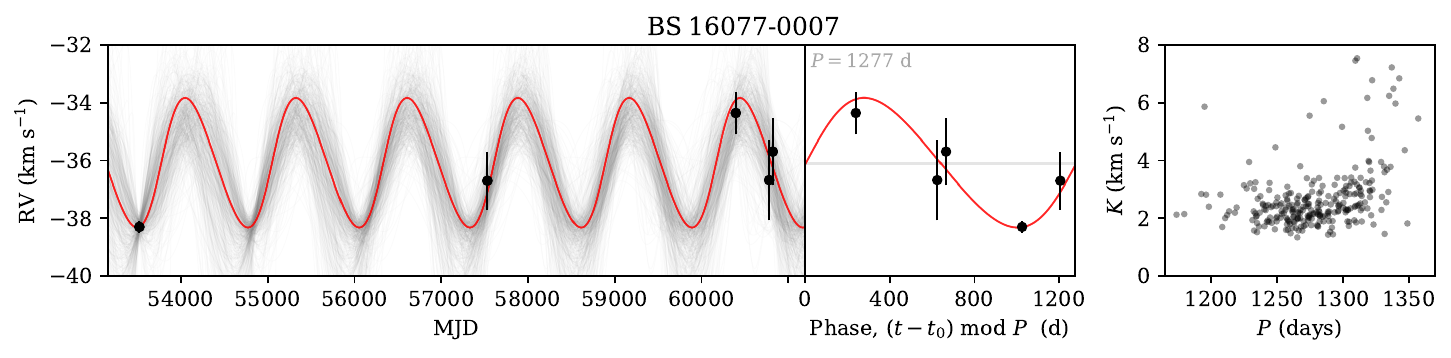}
\includegraphics[scale=0.73]{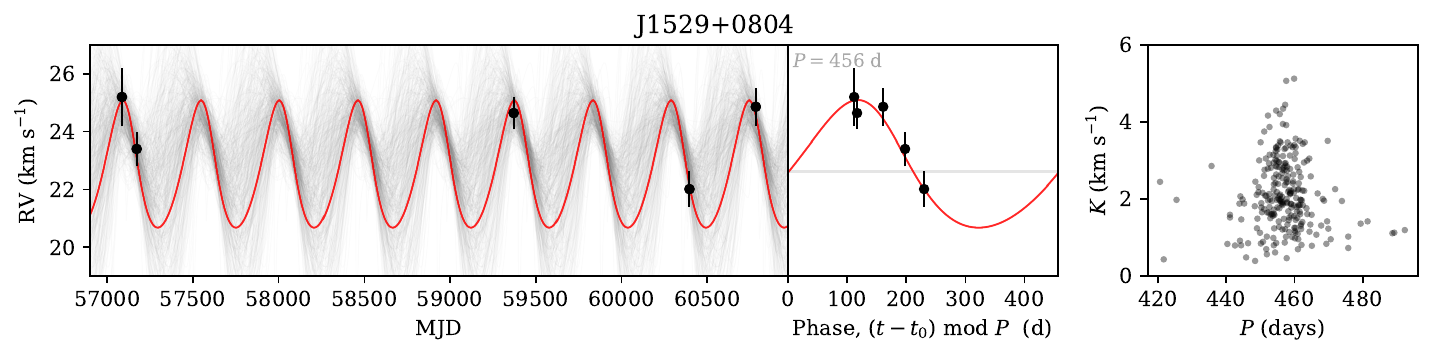}
\includegraphics[scale=0.73]{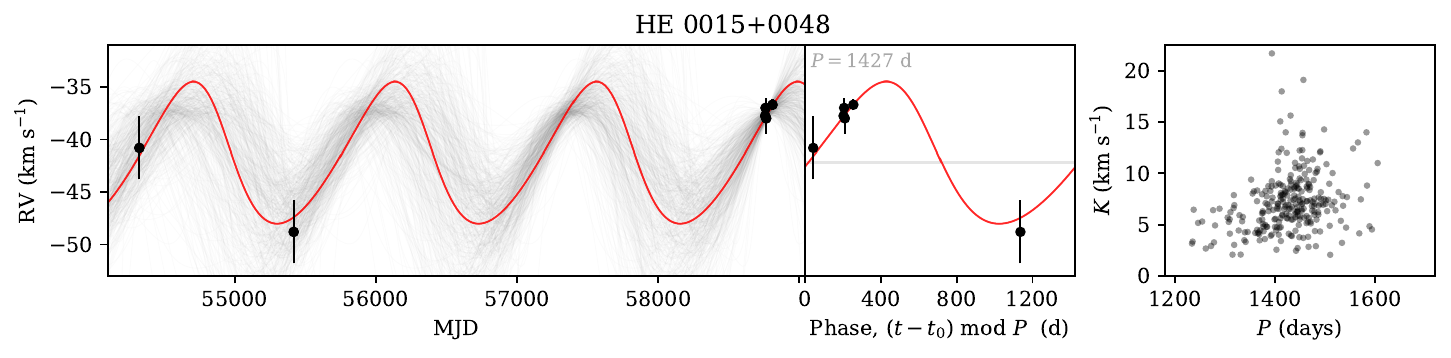}
\caption{\label{fig:orbits}Orbit fits from \texttt{The Joker} for BS~16077$-$0007, J1529$+$0804, and HE~0015$+$0048.
In the left panels, the prior samples that survived the rejection sampler are shown as thin gray curves, with the median-period orbit displayed in red. \textbf{The middle panels show the same RV data folded into a single phase.} The right panels display the periods and \textbf{RV semi-amplitudes} of the posterior distribution.}
\end{figure*}

\subsection{Orbital Parameters of Binaries}\label{sec:orbits}
\textbf{For three CEMP-no binaries with five or more RV measurements, we estimate a likely period ($P$) and RV semi-amplitude ($K$) using the Python-based Monte Carlo sampler \texttt{The Joker}. Based on a provided prior distribution, this code} constrains orbital parameters from very sparse or uncertain data, using a rejection sampling algorithm to generate a set of posterior orbits consistent with the provided RV data \citep{joker, joker_article}.

For these binaries, we generated $10^5$ prior samples and allowed for up to 512 posterior samples to be returned. \textbf{Based on the collected data, we set the reference scale of the $K$ prior variance ($\sigma_{K,0}$) to 30~\kms\ and the standard deviation of the prior systemic velocity ($\sigma_{v_0}$) to a deliberately broad value of 50~\kms. As our data are too limited to constrain an eccentricity, we kept the default beta distribution of eccentricities, with $a=0.867$ and $b=3.03$ \citep{joker_article}.} We initially ran the code for all stars with a period range (in days) of $10\le P_0\le10,000$, and based on the results, we narrowed our range to cover the strongest physical peak and ran \texttt{The Joker} again. The results are presented in Figure \ref{fig:orbits}, with the median-period posterior highlighted in red. \textbf{These are initial attempts at fitting orbits for the systems and are susceptible to change upon collection of additional RV data.}

Based on current RV data, we identify a likely period for J1529$+$0804 of $P=456\pm7$ days and a semi-amplitude of $K=2.00\pm0.86$~\kms. For BS~16077$-$0007, we find $P=1277\pm33$ days and $K=2.44\pm0.58$~\kms. For HE~0015$+$0048, we find $P=1427\pm66$ days and $K=6.88\pm2.58$~\kms. For all three stars, the eccentricities of the posteriors closely resemble the prior distribution, indicating that the rejection sampler was unable to constrain an eccentricity.

\subsection{CEMP-no Binary Frequency as a Function of $A(\mathrm{C})$}\label{sec:examining_binaries}

\begin{figure*}[ht!]
\centering
\includegraphics[scale=0.65]{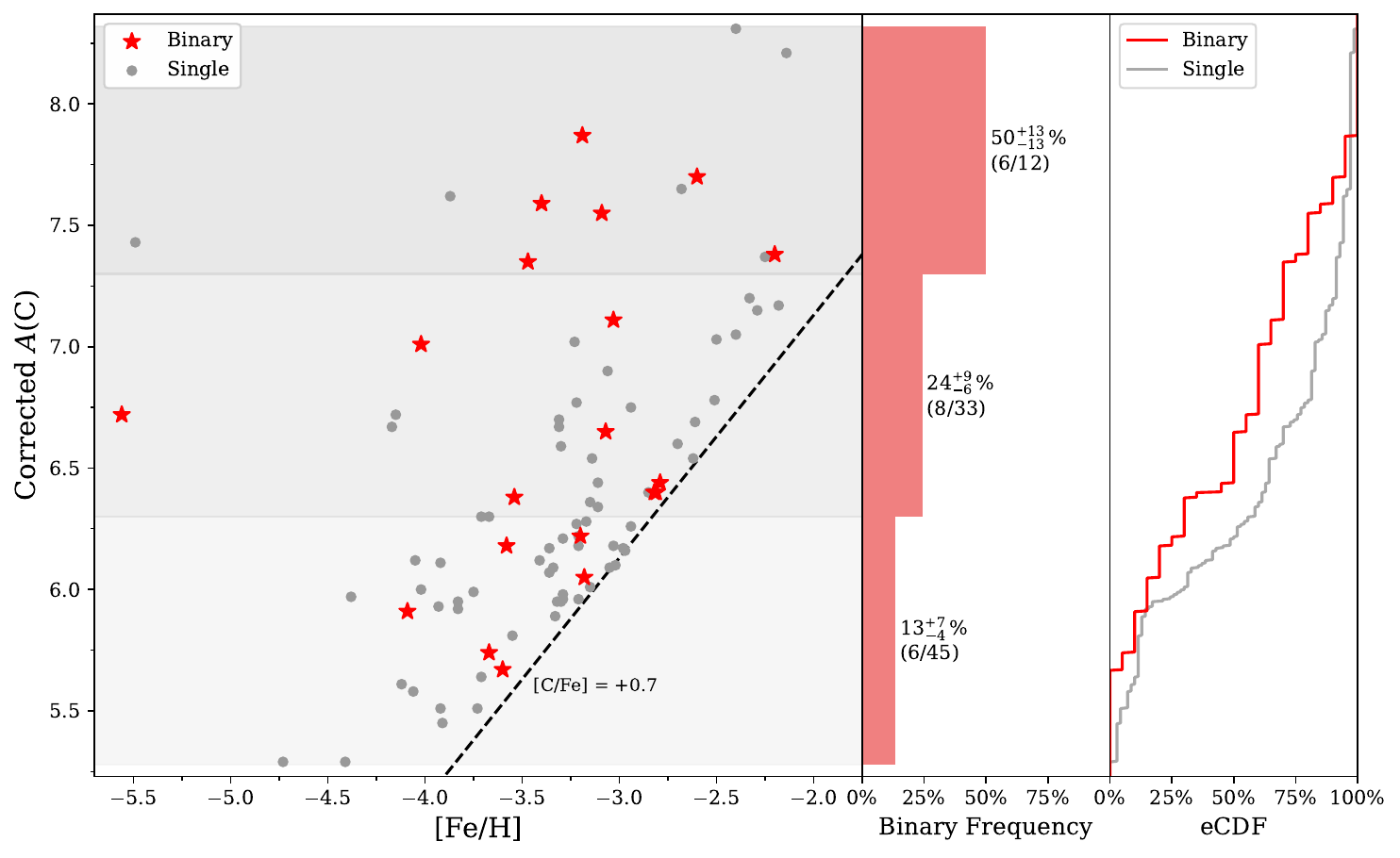}
\caption{\label{fig:YoonBeers_histogram}Absolute carbon abundances, corrected for the effects of stellar evolution, as a function of \FeH, for the 90 CEMP-no stars in our sample. The panels to the right show the binary frequency in three equally spaced \AC\ bins, as well as the empirical cumulative distribution functions (eCDFs) for binary and single CEMP-no stars.}
\end{figure*}

Our total sample of CEMP-no stars with constrained binary statuses includes 90 stars: 21 new statuses declared in this paper, and 69 statuses determined from literature data.
Due to the five-year monitoring period, we acknowledge that some stars declared single may actually be long-period binaries or systems with low inclination angles. However, face-on systems are fairly uncommon and \textbf{should not significantly impact our binary frequencies; even if our entire sample of 90 stars were binary, we should expect only 1 or 2 systems with $i\lesssim10^\circ$.}
Figure \ref{fig:YoonBeers_histogram} displays the full sample of 90 CEMP-no stars with a determined binary status in \AC-\FeH\ phase space, with binaries marked as red stars.

To investigate the potential association between binarity and \AC, we split our sample into three bands of equal width in carbon abundance: $5.3\le$ \AC\ $<6.3$, $6.3\le$ \AC\ $<7.3$, and $7.3\le$ \AC\ $<8.3$. 
In the low \AC\ band, 6 out of 45 stars are binary, giving a frequency of $13^{+7}_{-4}\%$. In the intermediate \AC\ band, 8 out of 33 stars are binary, giving a frequency of $24^{+9}_{-6}\%$. In the high \AC\ band, 6 out of 12 stars are binary, giving a frequency of $50^{+13}_{-13}\%$. The overall frequency for the entire sample is $22^{+5}_{-4}\%$.
Listed uncertainties are $1\sigma$ confidence intervals calculated using the Bayesian beta distribution method with a uniform prior. \textbf{Due to the small number of binaries per bin, this method is more reliable than the Gaussian interval or the continuity-corrected Wilson interval, which relies on a continuous normal approximation that becomes overly conservative at small sample sizes \citep{cameron2011,guan2012}.}

The combined CEMP-no binary frequency of $18^{+5}_{-4}\%$ among the two lower-carbon bands (\AC\ $<7.3$) is consistent with the general frequency of ${\sim}$15--20\% for metal-poor MW giants and solar-type stars \citep{carney2003,lodieu2025}. However, the binary frequency in the high \AC\ band differs by more than $2\sigma$ from low-carbon CEMP-no stars, demonstrating a clear trend in binarity as a function of \AC.

Additionally, we analyze the empirical cumulative distribution functions (eCDFs) of single and binary CEMP-no stars as a function of \AC, displayed in Figure \ref{fig:YoonBeers_histogram}.
The median values differ by 0.33\,dex (\AC\ $=6.22$ for single stars, 6.55 for binaries), while at the $80^{\rm{th}}$ percentile, this gap widens to 0.64\,dex (\AC\ $=6.77$ for single stars, 7.41 for binaries), driven by the prevalence of binaries in the high-carbon regime. \textbf{A two-sample Kolmogorov-Smirnov (KS) test yields a $p$-value of 0.0762, while a two-sample Anderson-Darling test, which is more sensitive to deviations near the tails of the distribution \citep{scholz1987}, yields a $p$-value of 0.0536. These tests fall slightly short of demonstrating significance at the $2\sigma$ level, likely due to the limited sample size.} 
\textbf{The uneven \AC\ distribution of CEMP-no binaries does not appear to reflect any underlying trend with \FeH. There is no clear correlation between \FeH\ and binary frequency, and a KS test between the binary and single eCDFs as a function of \FeH\ yields a statistically insignificant $p$-value of 0.833.} 

Finally, we note that most abundance analyses of these CEMP-no stars have used 1D local thermodynamic equilibrium (LTE) models, but 3D and non-LTE modeling is known to affect C and Fe abundances, possibly altering the CEMP labels, and changing the position in Figure \ref{fig:YoonBeers_histogram}, of some stars \citep{norris2019,lind2024}. \textbf{Thus, future 3D non-LTE analysis of CEMP-no stars is warranted to investigate potential impacts on the observed correlation between \AC\ and binary frequency.}

\section{Discussion}\label{sec:discussion}
Through extensive radial-velocity monitoring, we have investigated the binary frequency of CEMP-no stars, finding a significant increase for stars above \AC\ $=7.3$. Below, we discuss the potential implications of this correlation.

\subsection{Implications for Galactic Archaeology}\label{sec:implications}
For the low-carbon CEMP-no stars with \AC\ $<7.3$, our work confirms previous results from radial-velocity monitoring \citep{hansen2016a}, finding a binary frequency of $18^{+5}_{-4}$\%. This is consistent with the general frequency of \textbf{MW field giants and metal-poor solar-type stars \citep{carney2003,lodieu2025}, indicating that the abundances of low-carbon CEMP-no stars are most likely a record of the nucleosynthesis that occurred in the first generation of stars.}

The elevated binary fraction among high-carbon CEMP-no stars initially found by \citet{arentsen2019}—and now confirmed with this work—implies the existence of two formation channels for these stars: one coupled to the binary nature of the star and one likely similar to the formation channel(s) of low-carbon CEMP-no stars.
In recent work by \citet{lee2025}, a new group of CEMP-no stars was identified: Group~IV stars with \AC\ $>7.39$, \FeH\ $\le-3.1$. Group~IV occupies a morphologically distinct region of \AC-\FeH\ space that largely overlaps with our high-carbon bin of CEMP-no stars. \citet{lee2025} finds a binary frequency of ${\sim}$33\% (4/12) among their Group~IV candidates. We find a binary frequency of $50\pm13$\% among our high-carbon CEMP-no stars which cover a wider metallicity range, and specifically identify another binary candidate among their Group~IV stars (CS~22958$-$042). Our results thus support the elevated Group~IV binary frequency suggested by \citet{lee2025}.

The observed trend in binarity naturally beckons the question of whether the larger \AC\ values among CEMP-no binary stars are the result of mass transfer. 
The most likely companion star to have produced such large carbon abundances is a more massive star evolving through the AGB stage. However, current AGB nucleosynthesis models predict significant production of heavy elements like Sr and Ba via the slow neutron-capture process \citep[\sproc;][]{bisterzo2012,karakas2018}. 
This poses a severe challenge for Galactic chemical-evolution models, as the subsolar \AB{Ba}{Fe} of CEMP-no stars would imply that some metal-poor AGB stars produced lower $s$-process yields than predicted. This has been investigated for a few CEMP-no binary stars \citep{suda2004,lau2007}. However, there is no general consensus that some metal-poor AGB stars underproduce \sproc\ elements or that the third dredge-up can carry carbon to the stellar surface without also transporting \sproc\ material \citep{lugaro2023}.

\textbf{As an alternative to the mass-transfer explanation, it is also possible that the CEMP-no binary frequency depends more directly on \AC. For example, stellar natal clouds with large quantities of carbon may cool more efficiently than carbon-poor clouds \citep{glover2012}, potentially increasing binary formation as the cloud experiences more fragmentation. However, more modeling is needed to investigate the effect of carbon enhancement on the formation of binary systems.}

As the high-carbon CEMP-no binary frequency is much less than 100\%, an intrinsic enrichment source must also be invoked. \textbf{Faint population III supernovae are a commonly proposed formation channel; in the mixing-and-fallback model, the inner layers of the star ``fall back'' and form a black hole due to the low explosion energy, while the outer layers which contain more light elements like carbon are expelled \citep{umeda2003}.  Recent models suggest that these events cannot reproduce abundances above \AB{C}{H} $>-1$ (\AC\ $\gtrsim7.4$), corresponding to the lower envelope of the high-carbon CEMP-no stars \citep{komiya2020}, but \citet{lee2025} posits that inhomogeneous mixing of ejecta from multiple of these events may account for higher observed \AC\ abundances without substantial neutron-capture enrichment.}

\textbf{Stellar winds from rapidly rotating spinstars are another potential enrichment source for CEMP-no stars. Rotational mixing in the outer layers of massive stars can significantly boost $^{13}$C production, potentially explaining the lower $^{12}$C/$^{13}$C ratios of some CEMP-no stars \citep{maeder2015}.} \citet{choplin2026} modeled the ejecta of $20\,M_\odot$ spinstars and were able to reasonably reproduce the abundance pattern of the single CEMP-no star CS~29498$-$043 (\AC\ $=7.62$), demonstrating that spinstars are a plausible progenitor of high-carbon CEMP-no stars.

To fully constrain the origins of CEMP-no stars, more information is needed about their stellar abundances. \citet{yoon2016} found different behavior in the \ANa-\AC\ and \AMg-\AC\ morphology for Group~II and Group~III CEMP-no stars, and \citet{hartwig2018} established a significant difference in \AB{Mg}{C} between mono-enriched and multi-enriched second-generation stars. \textbf{Additionally, more $^{12}$C/$^{13}$C isotopic ratios of CEMP-no stars will help us to probe the level of mixing in their progenitors.} This isotopic ratio was examined by \citet{molaro2023}, who found a potential shift in the $^{12}$C/$^{13}$C ratio at around \FeH\ $=-4$.

\subsection{The Nature of the Unseen Companion in Binary CEMP-no Systems}\label{sec:nature_of_companion}

For the \textbf{three} binary systems with newly constrained orbital parameters (J1529$+$0804, BS~16077$-$0007, and HE~0015$+$0048), we can investigate the possibility of AGB mass transfer by placing constraints on the nature of the unseen companion, \textbf{since AGB stars evolve into white dwarfs (WDs) with a known mass range of ${\sim}$0.4--1.4\,$M_\odot$ \citep{kepler2007}}. For \textit{all} high-resolution spectra of the binaries in our sample, we only observe a single set of absorption lines with no signs of blending or asymmetry, implying a large luminosity difference \textbf{consistent with, but not exclusive to,} a WD companion. Double-lined spectroscopic binaries \textbf{may be undetected if observations are taken near conjunction, though this becomes highly unlikely as the number of observations increases}.

Assuming nearly circular orbits \textbf{and propagating the uncertainties on $P$ and $K$, we calculate binary mass functions of $f\simeq0.00038^{+0.00073}_{-0.00031}\,M_\odot$ for J1529$+$0804, $f\simeq0.0019^{+0.0017}_{-0.0011}\,M_\odot$ for BS~16077$-$0007, and $f\simeq0.048^{+0.077}_{-0.036}\,M_\odot$ for HE~0015$+$0048. The inclination angles ($i$) of these systems are unknown, so we adopt the statistical average $\langle\sin^3{i}\rangle=3\pi/16\simeq0.589$ and assume a primary mass of $0.8\,M_\odot$, typical of metal-poor MW giants \citep{carney2003}. With these assumptions, we estimate secondary masses ($M_2$) of $0.08^{+0.04}_{-0.04}$, $0.14^{+0.04}_{-0.03}$, and $0.52^{+0.30}_{-0.23}\,M_\odot$, respectively.}
For BS~16077$-$0007 or J1529$+$0804 to have a WD companion, their orbital inclination angles would both need to be very small \textbf{($i\lesssim20^\circ$), which is unlikely to occur twice within a sample of 20 binaries. The secondary mass of HE~0015$+$0048 falls within the WD mass range; however, our $M_2$ estimate depends on $P$ and $K$ which are uncertain due to clustered observations, as well as assumed values for $i$ and the eccentricity ($e$), which cannot be constrained.}
Further observations of these three stars are warranted to investigate the likelihood of a WD companion.

Previous studies have also derived orbits for other CEMP-no binaries, allowing us to similarly estimate $M_2$ using reported values of $P$, $K$, and $e$ (if known), \textbf{and the assumption of $\sin^3{i}=\langle\sin^3{i}\rangle\simeq0.589$.}
\begin{itemize}
    \item \citet{hansen2016a} determined that three CEMP-no binaries have orbits that could plausibly support Roche-lobe overflow from an AGB-star companion. Using their derived orbits, we find $M_2\simeq0.59\pm0.02\,M_\odot$, $M_2\simeq0.49\pm0.01\,M_\odot$, and $M_2\simeq0.52\pm0.02\,M_\odot$ for HE~0219$-$1739, HE~1150$-$0428, and CS~22957$-$027, respectively.
    \item \citet{bandyopadhyay2018} reports $P=116$ days and $K\simeq55$~\kms\ for J1341$+$4741, giving a rough estimate of $M_2\simeq4.6\,M_\odot$. Tighter orbital constraints might lower this estimate, but current RV data suggest a WD companion is unlikely.
    \item \citet{arentsen2019} reports $P=32$ days and $K=7.5$~\kms\ for J0140$+$2344, giving $M_2\simeq0.13\,M_\odot$. This study also finds $P\simeq1600$ days and $K\simeq8$~\kms\ for J1422$+$0031, giving $M_2\simeq0.68\,M_\odot$.
    \item \citet{caffau2025} reports $P=10560^{+786}_{-518}$ days, $K=1.67^{+0.08}_{-0.07}$~\kms, and $e=0.24^{+0.08}_{-0.08}$ for HE~0107$-$5240, giving $M_2\simeq0.20\pm0.01\,M_\odot$.
\end{itemize}
            
\textbf{Of the ten CEMP-no binary systems discussed above, all three high-carbon CEMP-no binaries (\AC\ $\ge7.3$) have $M_2$ estimates within the WD range, while five of the seven stars with \AC\ $<7.3$ have $M_2$ estimates that are inconsistent with a WD. Four of these (J1529$+$0804, BS~16077$-$0007, J0140$+$2344, and HE~0107$-$5240) might be better explained by a less massive binary companion, while the $M_2$ estimate for J1341$+$4741 suggests a potential compact object companion or a higher-multiplicity system. However, we reemphasize (for all stars) that these $M_2$ estimates are very loosely constrained and may significantly change with future RV measurements.}

\subsection{Sr and Ba Abundances in CEMP-no Stars}
In addition to the orbital parameters of binaries, we can also seek information from the abundances of elements detected in the atmospheres of CEMP-no stars. In particular, neutron-capture elements hold clues about the progenitors of the star.

\textbf{Sr and Ba have been detected even in the most metal-poor stars \citep{roederer2013}, highlighting that a production channel must have been active in the very early universe.} These elements can be produced via the main $s$-process during thermal pulses in AGB stars, as well as the weak $s$-process in massive stars \citep{lugaro2023}. In particular, stellar rotation is known to boost $s$-process production in massive stars \citep{frischknecht2016}.  However, the \AB{Sr}{Ba} ratio likely differs between the two production sites, with the weak $s$-process generally producing a larger \AB{Sr}{Ba} ratio compared to the main $s$-process \citep{cristallo2011,frischknecht2012}. Thus, the \AB{Sr}{Ba} ratio serves as a diagnostic tool to assess which astrophysical site is most likely responsible for the chemical signature of CEMP stars. \citet{hansen2019} used this abundance criterion to separate CEMP stars enriched by AGB stars from those enriched by massive stars. If a CEMP-no star truly lacks enrichment in the main $s$-process, and its Ba instead arises from the ejecta of a massive star, we should observe a very large \AB{Sr}{Ba} ratio. If the star is instead enriched by an AGB companion, we should expect $-0.5\lesssim$ \AB{Sr}{Ba} $\lesssim+0.75$ \citep{hansen2019}.

\begin{figure}[]
\centering
\includegraphics[scale=0.61]{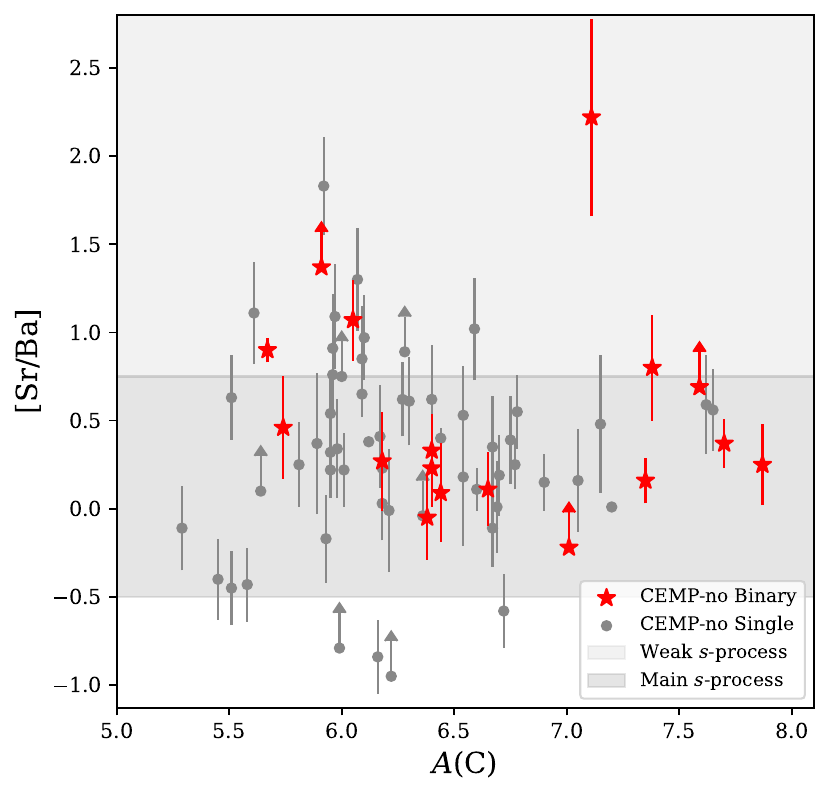}
\caption{\label{fig:SrBa_plots} \AB{Sr}{Ba}, as a function of absolute carbon abundance, for the CEMP-no sample. Shaded regions indicate potential enrichment via the main \sproc\ (AGB stars) and weak \sproc\ (massive stars) from \citet{hansen2019}.}
\end{figure}

By definition, all stars in our sample have a previously derived Ba abundance or upper limit, and a majority (73/90) have a reported Sr abundance as well, allowing us to analyze the \AB{Sr}{Ba} ratio for most of our sample. Figure \ref{fig:SrBa_plots} shows \AB{Sr}{Ba} values for our CEMP-no sample plotted as a function of \AC; color coding is the same as Figure \ref{fig:YoonBeers_histogram}. The shaded areas show \AB{Sr}{Ba} abundances compatible with enrichment by AGB stars and massive stars as defined in \citet{hansen2019}.

Interestingly, a large majority of CEMP-no stars with a reported Sr abundance, including all of the high-carbon binaries, are consistent with the \AB{Sr}{Ba} range typically produced via the main $s$-process in AGB stars. However, a substantial amount of single CEMP-no stars also have \AB{Sr}{Ba} ratios in this range. It is very unlikely that the Sr and Ba present in the most metal-poor single CEMP-no stars were created in AGB stars, as AGB stars would not have had time to pollute the interstellar medium at the time of star formation. Hence, \AB{Sr}{Ba} does not necessarily provide a clear picture of the progenitors of the CEMP-no stars.

\begin{figure*}[ht!]
\centering
\includegraphics[scale=0.6,trim={0 0 0 0}]{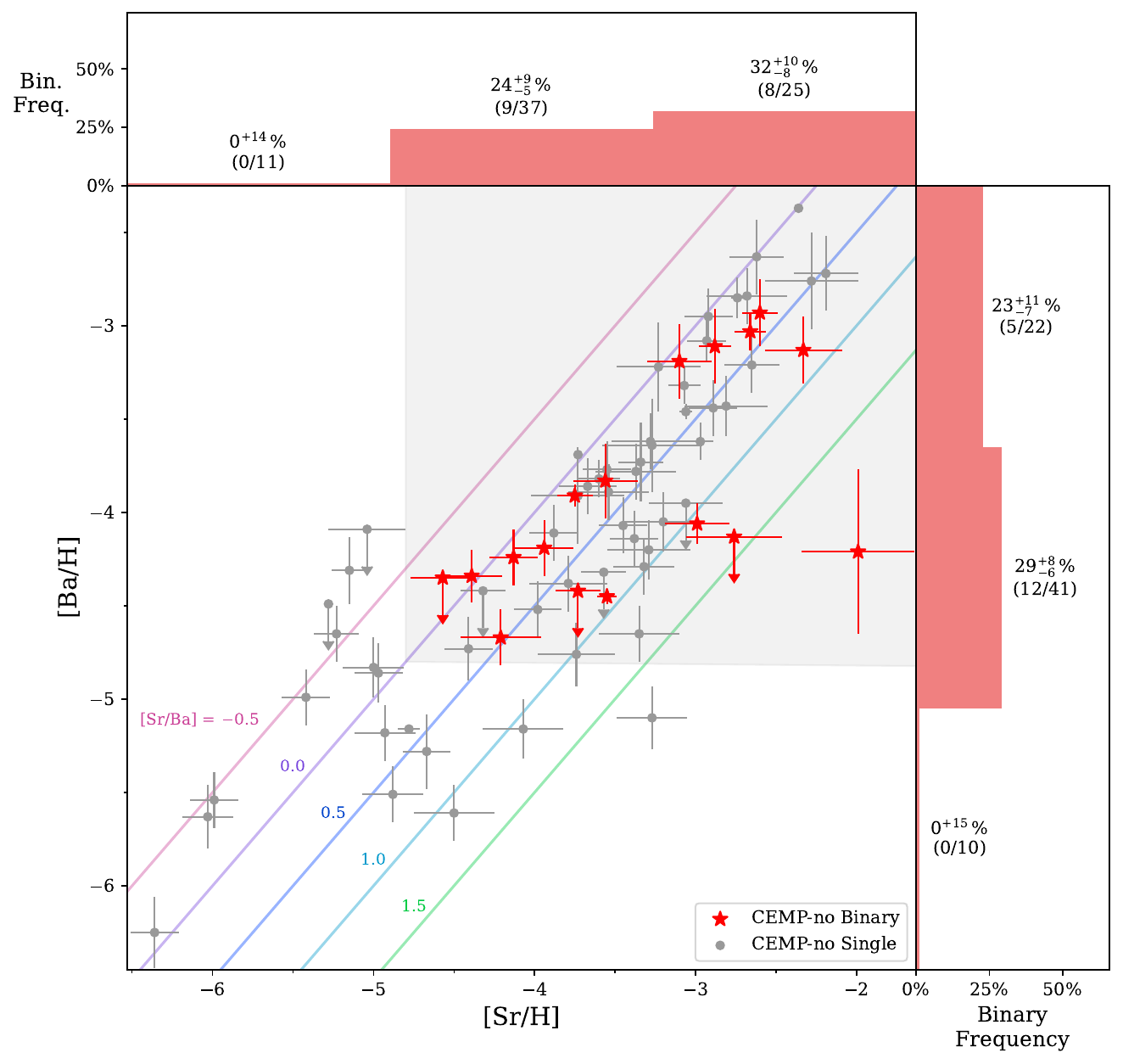}
\caption{\label{fig:Ba_vs_Sr} \AB{Ba}{H}, as a function of \AB{Sr}{H}, for stars in our sample with a known \AB{Sr}{Ba} abundance or lower limit. The gray shaded region indicates where binaries are found, with binary frequencies given in the histograms. Diagonal lines of constant \AB{Sr}{Ba} are also shown. }
\end{figure*}

The individual Sr and Ba abundances, on the other hand, might provide some clues. Figure \ref{fig:Ba_vs_Sr} shows the distribution of CEMP-no binaries with a known \AB{Sr}{Ba} abundance or lower limit, as a function of \AB{Ba}{H} and \AB{Sr}{H}, removing iron from both axes. Interestingly, all binaries in this plot have abundances higher than \AB{Ba}{H} $=$  \AB{Sr}{H} $=-4.8$. This is illustrated by the shaded region of Figure \ref{fig:Ba_vs_Sr}, inside which we determine an overall binary frequency of ${\sim}30\%$, but no clear trend with Ba or Sr. The stars outside this region, due to the lack of observed binaries and very low metallicity, can be more confidently labeled as true second-generation stars.

\section{Summary}\label{sec:summary}

We present the results of radial-velocity monitoring of 30 CEMP-no stars over a period of five years, in order to investigate a potential correlation between the binary frequency of CEMP-no stars and their absolute carbon abundances, as suggested by \citet{arentsen2019}.
In our monitoring sample, we identify four new binaries and confirm the statuses of two previously declared binaries. Twenty-two CEMP-no stars are declared single, while two stars remain uncertain until more radial-velocity measurements are published. 

Including results from the literature, we compiled a total of 90 CEMP-no binary statuses. We derived a binary fraction of $50^{+13}_{-13}\%$ for high-carbon CEMP-no stars (\AC\ $\ge7.3$), compared to $18^{+5}_{-4}\%$ for low-carbon CEMP-no stars (\AC\ $<7.3$); the latter aligns with the typical binary frequency of metal-poor Milky Way halo stars. Our observations confirm a link between binarity and \AC\ for CEMP-no stars, suggesting that a binary-related formation channel could be responsible for the abundance signature seen in roughly half of high-carbon CEMP-no stars, while the other half (and low-carbon CEMP-no stars) likely preserve a record of the nucleosynthesis within the first generation of stars. 

To further investigate the possibility of a mass-transfer-related formation channel, we explore the nature of the unseen companion in our confirmed binary systems. From this analysis, we found that a majority of high-carbon CEMP-no binary systems could plausibly host the remnants of former AGB stars. Further evidence for AGB mass transfer could be detectable in the \AB{Sr}{Ba} ratios of CEMP-no stars, but we find no correlation between binary frequency and \AB{Sr}{Ba} among our current sample. Thus, a combination of chemical analysis and further radial-velocity monitoring is needed to better constrain orbits and understand the nature of CEMP-no stars.

For this work, the upcoming fourth data release of the \textit{Gaia} mission\footnote{\url{https://www.cosmos.esa.int/web/gaia/dr4}} is promising, as it will provide individual-epoch RVs which allow us to quickly and reliably determine the binarity, and potentially orbital parameters, of many brighter CEMP-no stars. Combining this information with abundance results from the 4MOST \citep{dejong2019} and WEAVE \citep{shoko2024} surveys will \textbf{allow us to further constrain the elusive origins of CEMP-no stars and better understand early trends in Galactic chemical evolution.}

\begin{acknowledgments}

We thank the anonymous referee for their constructive feedback that has helped to improve this manuscript. We are also grateful to Dr. Camilla J. Hansen for valuable discussions regarding Sr and Ba abundances which led to the analysis in Section 5.3.

This work has made use of data from the European Space Agency (ESA) mission {\it Gaia}, processed by the {\it Gaia} Data Processing and Analysis Consortium (DPAC,
\url{https://www.cosmos.esa.int/web/gaia/dpac/consortium}). Funding for the DPAC has been provided by national institutions, in particular the institutions participating in the {\it Gaia} Multilateral Agreement. NOIRLab IRAF is distributed by the Community Science and Data Center at NOIRLab, which is managed by the Association of Universities for Research in Astronomy (AURA) under a cooperative agreement with the U.S. National Science Foundation (NSF).

This work is based on observations collected at the McDonald Observatory of the University of Texas at Austin; the European Southern Observatory telescopes under program IDs 0104.A-9008(A) and 170.D-0010; the Gemini Remote Access to CFHT ESPaDOnS Spectrograph (GRACES) under program ID GN-2019B-Q-219; and the Network of Robotic Echelle Spectrographs (NRES) from the Las Cumbres Observatory global telescope network, which was developed through financial support from NSF MRI (AST-1229720) and ATI (1407666 \& 1508464) grants. This research has made use of the services of the SIMBAD database, operated at CDS, Strasbourg, France, as well as the Stellar Abundances for Galactic Archaeology (SAGA) database.  

This material is based upon work supported by the NSF Graduate Research Fellowship Program under Grant No.\@ DGE-2139772, with additional travel support provided by a Doctoral Student Research, Travel, and Professional Development Award from the College of Arts and Sciences at Texas A\&M University. Any opinions, findings, and conclusions or recommendations expressed in this material are those of the authors and do not necessarily reflect the views of the NSF. This research benefited from support from the George P. and Cynthia Woods Mitchell Institute for Fundamental Physics and Astronomy, the Dr.\@ Dionel Avilés ’53 and Dr.\@ James Johnson ’67 Graduate Fellowship Program, and Charles R. ’62 and Judith G. Munnerlyn at Texas A\&M University.
This work also benefited from discussions at the Third Frontiers Summer School and the 2025 Frontiers in Nuclear Astrophysics meeting at Ohio University, supported by the International Research Network for Nuclear Astrophysics (IReNA) under NSF Grant OISE-1927130.

J.L.M. thanks Dr. Ersen Arseven for generous support of this research. 
T.T.H.\ acknowledges support from the Swedish Research Council (VR 2021-05556).
The work of V.M.P. is supported by NOIRLab, which is managed by the Association of Universities for Research in Astronomy (AURA) under a cooperative agreement with the U.S. NSF. 
T.C.B. acknowledges partial support from grants PHY 14-30152; Physics Frontier Center/JINA Center for the Evolution of the Elements (JINA-CEE), and OISE-1927130; IReNA, awarded by the NSF, and DE-SC0023128; the Center for Nuclear Astrophysics Across Messengers (CeNAM), awarded by the U.S. Department of Energy, Office of Science, Office of Nuclear Physics.
%

\end{acknowledgments}

\facilities{Smith, LCOGT, Max Planck:2.2m, 
VLT:Kueyen, Gemini:Gillett
}
\software{\texttt{matplotlib}~\citep{matplotlib},
NOIRLab \texttt{IRAF}~\citep{iraf1,iraf2,iraf3},
astropy \citep{2013A&A...558A..33A,2018AJ....156..123A,2022ApJ...935..167A},  
\texttt{The Joker} \citep{joker, joker_article}
MOOG \citep{sneden1973,sobeck2011},
linemake \citep{placco2021,linemake},
SMHR \citep{casey2025}
}
\bibliography{ref.bib}{}
\bibliographystyle{aasjournalv7.1}

\appendix\restartappendixnumbering

\section{High-Resolution Heliocentric Radial-Velocity Measurements}\label{sec:appendix_velocities}

In Appendix Table \ref{tab:appendix_rvs}, we present our measured RVs, as well as RVs from high-resolution observations reported in the literature, for the 30 CEMP-no targets discussed above. In Table \ref{tab:literature_stars}, we provide details of the 62 stars collected from past literature that are shown in Figure \ref{fig:YoonBeers_histogram} and used for statistical analysis of the CEMP-no binary frequency.

\startlongtable
\begin{deluxetable*}{llccrrc}
\tablecaption{\label{tab:appendix_rvs} RVs of 30 CEMP-no Stars from our Observations and the Literature}
\tablehead{\colhead{Star Name} & \colhead{HJD} & \colhead{S/N at 5100~\AA} & $N{}^a$ & \colhead{$v_r$} & \colhead{$\sigma_{v_r}$} & \colhead{Reference}  \\
{} & {} & {} & {} & (\kms) & (\kms) & {}
} 
\startdata
HE~0015$+$0048   &  2458810.51018   & 16 &    17  & $-36.69$          &  $0.57$            &  This work (FEROS)     \\
{}             &  2458765.50692   &  5 &    17  & $-37.99$          &  $1.48$            &  This work (FEROS)     \\
{}             &  2458761.50663   &  7 &    11  & $-36.99$          &  $0.96$            &  This work (FEROS)     \\
{}             &  2458758.50467   &  9 &     6  & $-37.75$          &  $0.40$            &  This work (FEROS)     \\
{}             &  2455416         & {} &    {}  & $-48.8$            &  $3^b$              &  \citet{hollek2011}     \\
{}             &  2454322         & {} &    {}  & $-40.8$            &  $3^b$              &  \citet{hollek2011}     \\
{}             &  …               & {} &    {}  & $\mathit{-49.98}$  & $\mathit{1.44}$     &  \textit{Gaia DR3}    \\ \hline
CS~22958$-$042   &  2458863.52869   & 13 &     9  & $175.25$         &  $1.97$         &  This work (FEROS)      \\
{}             &  2458810.51921   & 13 &     6  & $178.26$         &  $1.80$         &  This work (FEROS)      \\
{}             &  2458761.51387   &  7 &     5  & $176.11$         &  $1.77$         &  This work (FEROS)      \\
{}             &  2452658.550     & {} &    {}  & $165.6$           &  $0.8$         &  \citet{roederer2014}     \\
{}             &  2452656.547     & {} &    {}  & $165.7$           &  $0.8$         &  \citet{roederer2014}     \\
{}             &  2452655.561     & {} &    {}  & $165.2$           &  $0.8$         &  \citet{roederer2014}     \\
{}             &  2452218.23606   & {} &    {}  & $162.51$          &  $3^b$         &  \citet{sivarani2006}     \\
{}             &  2452218.19002   & {} &    {}  & $165.21$          &  $3^b$         &  \citet{sivarani2006}     \\  \hline
\enddata
\tablerefs{\textit{Gaia DR2}: \citet{Gaia_collab,Gaia_DR2,Gaia_DR2_RVs}\\
\textit{Gaia DR3}: \citet{Gaia_collab,Gaia_DR3,Gaia_DR3_RVs}
} 
\tablecomments{Appendix Table \ref{tab:appendix_rvs} is available in its entirety in machine-readable format. Measurements for two stars are shown here to illustrate its form and content.
Gaia RVs are included for reference when available, but were not used for determination of binary statuses.}
\tablenotetext{a}{For our new observations, N is the number of echelle orders used in the cross-correlation to determine an RV.}
\tablenotetext{b}{An uncertainty of 3~\kms\ is assumed when no value is provided in the reference paper.}
\end{deluxetable*}

\startlongtable
\begin{deluxetable}{lrrrrrrcc}
\tablecaption{\label{tab:literature_stars} Binary Statuses of Literature CEMP-no Stars Displayed in Figure \ref{fig:YoonBeers_histogram}}
\tablehead{\colhead{Star Name} & \colhead{Gaia DR3 ID} & \colhead{RA} & \colhead{DEC} & \colhead{\FeH} & \colhead{\AC} & \colhead{\BaFe} & \colhead{Binary?} & \colhead{Ref.}  \\
{} & {} & (hh mm ss) & (dd mm ss) & {} & {} & {} & {} & {} 
} 
\startdata
HE~0020$-$1741       &  2367173119271988480   &  00 22 44.00  & $-$17 24 28.0 &  $-4.05$ &  6.12  &  $-1.11$  &  No     &   1, 2                     \\
J0039$-$6849         &  4702817360364840704   &  00 39 46.91  & $-$68 49 57.4 &  $-2.62$ &  6.54  &  $-0.10$  &  No     &   3, 4                     \\
CS~22166$-$016       &  2372010867355028608   &  00 58 23.84  & $-$14 47 06.9 &  $-2.40$ &  7.05  &  $-0.44$  &  No     &   5, 6, 7                  \\
HE~0102$-$0633       &  2476709216409802880   &  01 05 25.60  & $-$06 17 36.0 &  $-3.15$ &  6.36  & $<-0.54$  &  No     &   8                        \\
HE~0107$-$5240       &  4927204800008334464   &  01 09 29.16  & $-$52 24 34.2 &  $-5.56$ &  6.72  & $<-0.48$  & Yes     &   1, 7, 9                  \\
CS~22953$-$037       &  4716937597925985152   &  01 25 06.65  & $-$59 16 00.7 &  $-3.21$ &  6.22  & $<-0.88$  &  No     &   1, 5                     \\
HE~0132$-$2439       &  5038842338744610816   &  01 34 58.80  & $-$24 24 18.0 &  $-3.60$ &  5.67  &  $-0.85$  & Yes     &   1, 8                     \\
J0140$+$2344         &   290930261314166528   &  01 40 36.22  & $+$23 44 58.1 &  $-4.09$ &  5.91  & $<-0.04$  & Yes     &   1, 7                     \\
CS~22958$-$083       &  4743992799514732800   &  02 15 42.73  & $-$53 59 56.3 &  $-3.05$ &  6.09  &  $-1.00$  &  No     &   1, 5                     \\
HE~0219$-$1739       &  5143740592756919808   &  02 21 41.00  & $-$17 25 37.0 &  $-3.09$ &  7.55  & $<-1.39$  & Yes     &   1, 6                     \\
BD$+$44~493          &   341511064663637376   &  02 26 49.74  & $+$44 57 46.5 &  $-3.83$ &  5.95  &  $-0.90$  &  No     &   1, 5, 10                 \\
CS~22189$-$009       &  5170309947645049728   &  02 41 42.38  & $-$13 28 10.5 &  $-3.92$ &  5.51  &  $-1.59$  &  No     &   1, 5                     \\
HE~0324$-$0152       &  3268028903151246720   &  03 26 53.80  & $+$02 02 28.0 &  $-3.32$ &  5.95  &  $-1.20$  &  No     &   1, 11                    \\
HE~0405$-$0526       &  3197521276912546432   &  04 07 47.00  & $-$05 18 11.0 &  $-2.18$ &  7.17  &  $-0.22$  &  No     &   1, 6                     \\
HE~0432$-$1005       &  3185294604533440640   &  04 35 01.31  & $-$09 59 36.3 &  $-3.21$ &  6.18  &  $-0.90$  &  No     &   1, 11                    \\
HE~0557$-$4840       &  4794791782906532608   &  05 58 39.26  & $-$48 39 56.8 &  $-4.73$ &  5.29  &  $<0.00$  &  No     &   1, 7, 12                 \\
J0815$+$4729         &   931227322991970560   &  08 15 54.27  & $+$47 29 47.6 &  $-5.49$ &  7.43  & $<-3.58$  &  No     &   7, 13                    \\
HD~237846            &  1049376272667191552   &  09 52 38.69  & $+$57 54 58.4 &  $-3.21$ &  5.96  &  $-0.99$  &  No     &   1, 5                     \\
HE~1116$-$0634       &  3783989199935482624   &  11 18 35.89  & $-$06 50 45.1 &  $-3.73$ &  5.51  &  $-1.81$  &  No     &   1, 11                    \\
HE~1124$-$2335       &  3534825915028932608   &  11 27 26.95  & $-$23 52 05.6 &  $-3.36$ &  6.07  &  $-1.29$  &  No     &   1, 5                     \\
HE~1133$-$0555       &  3593627144045992832   &  11 36 12.00  & $+$06 11 43.0 &  $-2.40$ &  8.31  &  $-0.58$  &  No     &   1, 6                     \\
HE~1141$-$0610       &  3592672424356160896   &  11 44 22.60  & $-$06 26 59.0 &  $-2.33$ &  7.20  &  $-0.04$  &  No     &   8                        \\
HE~1150$-$0428       &  3599270524915072000   &  11 53 06.60  & $-$04 45 03.4 &  $-3.47$ &  7.35  &  $-0.44$  & Yes     &   1, 14                    \\
HE~1201$-$1512       &  3568795670365205888   &  12 03 37.10  & $-$15 29 32.0 &  $-3.92$ &  6.11  &  $<-0.34$ &  No     &   1, 14                    \\
BS~16076$-$006       &  3954415903126795136   &  12 48 22.75  & $+$20 56 44.0 &  $-3.36$ &  6.17  &  $<-1.06$ &  No     &   15, 16                   \\
HE~1300$+$0157       &  3691867477195410944   &  13 02 56.24  & $+$01 41 52.1 &  $-3.75$ &  5.99  &  $<-0.74$ &  No     &   1, 14                    \\
BS~16929$-$005       &  1467915131946458496   &  13 03 29.47  & $+$33 51 09.1 &  $-3.34$ &  6.09  &  $-0.28$  &  No     &   1, 14, 17                \\
HE~1300$-$0641       &  3629020217185444352   &  13 03 34.14  & $-$06 57 20.8 &  $-3.14$ &  6.54  &  $-0.77$  &  No     &   1, 18                    \\
HE~1302$-$0954       &  3623099450149184768   &  13 04 58.00  & $-$10 10 11.0 &  $-2.25$ &  7.37  &  $<-0.53$ &  No     &   1, 6                     \\
BS~16033$-$081       &  1445684243782865792   &  13 19 12.46  & $+$22 27 57.6 &  $-2.20$ &  7.38  &  $-0.93$  & Yes     &   19                       \\
G~64$-$12            &  3662741860852094208   &  13 40 02.50  & $-$00 02 18.8 &  $-3.29$ &  6.21  &  $+0.07$  &  No     &   7, 20, 21                \\
G~64$-$37            &  3643857920443831168   &  13 40 02.50  & $-$00 02 18.8 &  $-3.11$ &  6.44  &  $-0.35$  &  No     &   20, 21                   \\
J1341$+$4741         &  1552204895923527040   &  13 41 44.61  & $+$47 41 28.9 &  $-3.20$ &  6.22  &  $-0.73$  & Yes     &   7                        \\
HE~1410$+$0213       &  3667206118578976896   &  14 13 06.56  & $+$01 59 21.9 &  $-2.14$ &  8.21  &  $-0.26$  &  No     &   1, 8                     \\
J1422$+$0031         &  3654012700600277888   &  14 22 37.43  & $+$00 31 05.2 &  $-3.03$ &  7.11  &  $-1.18$  & Yes     &   1, 7, 22                 \\
HD~126587            &  6273496177641286272   &  14 27 00.36  & $-$22 14 39.0 &  $-3.29$ &  5.98  &  $-0.33$  &  No     &   1, 5                     \\
HE~1506$-$0113       &  4418328965879435776   &  15 09 14.30  & $-$01 24 57.0 &  $-3.54$ &  6.38  &  $-0.80$  & Yes     &   1, 14                    \\
CS~22878$-$027       &  4446665716995846400   &  16 37 35.84  & $+$10 22 07.8 &  $-2.51$ &  6.78  &  $-0.93$  &  No     &   5, 7                     \\
J1645$+$4357         &  1357725650023190784   &  16 45 14.95  & $+$43 57 12.1 &  $-2.97$ &  6.16  &  $-1.34$  &  No     &   23, 24, 25               \\
J1738$-$1457         &  4125223698812069632   &  17 38 23.36  & $-$14 57 01.0 &  $-3.58$ &  6.18  &  $-0.25$  & Yes     &   1, 26, 27                \\
J1917$-$5440         &  6643680508396027776   &  19 17 55.86  & $-$54 40 14.8 &  $-2.81$ &  6.40  &  $-0.30$  & Yes     &   3, 4                     \\
CS~22891$-$200       &  6445220927325014016   &  19 35 19.06  & $-$61 42 24.5 &  $-4.06$ &  5.58  &  $-0.93$  &  No     &   1, 5                     \\
CS~22896$-$110       &  6645203744317443968   &  19 35 48.02  & $-$53 26 16.3 &  $-2.85$ &  6.40  &  $-0.58$  &  No     &   1, 5                     \\
CS~22885$-$096       &  6692925538259931136   &  20 20 51.16  & $-$39 53 30.3 &  $-4.41$ &  5.29  &  $-1.84$  &  No     &   1, 5                     \\
CS~22943$-$201       &  6676149533440551936   &  20 36 28.09  & $-$43 49 28.9 &  $-2.68$ &  7.65  &  $-0.53$  &  No     &   1, 5                     \\
J2037$-$1221         &  6900062671257456000   &  20 37 06.35  & $-$12 21 25.1 &  $-2.61$ &  6.69  &  $-0.02$  &  No     &   3, 28                    \\
CS~22897$-$008       &  6449369934453211264   &  21 03 11.87  & $-$65 05 08.8 &  $-3.83$ &  5.92  &  $-1.27$  &  No     &   1, 5                     \\
HE~2123$-$0329       &  2684404076678702208   &  21 26 08.96  & $-$03 16 58.8 &  $-3.22$ &  6.27  &  $-0.85$  &  No     &   1, 11                    \\
HE~2138$-$0314       &  2674056091713363840   &  21 40 41.59  & $-$03 01 17.1 &  $-3.29$ &  5.96  &  $-0.85$  &  No     &   1, 11                    \\
HE~2139$-$5432       &  6461736966363075200   &  21 42 42.50  & $-$54 18 43.0 &  $-4.02$ &  7.01  &  $<-0.33$ & Yes     &   1, 7                     \\
CS~22956$-$050       &  6399358510623784192   &  21 58 05.82  & $-$65 13 27.2 &  $-3.67$ &  5.74  &  $-1.00$  & Yes     &   1, 5, 26                 \\
CS~22965$-$054       &  2676443097097288704   &  22 06 30.30  & $-$02 32 34.5 &  $-3.17$ &  6.28  &  $<-0.78$ &  No     &   1, 5                     \\
CS~22960$-$048       &  6567131271918958720   &  22 17 01.57  & $-$45 12 16.9 &  $-3.91$ &  5.45  &  $-1.72$  &  No     &   1, 5                     \\
J2217$+$2104         &  1778804140643594240   &  22 17 50.59  & $+$21 04 37.2 &  $-3.93$ &  5.93  &  $-0.90$  &  No     &   23, 25, 29               \\
CD$-$24~17504        &  2383484851010749568   &  23 07 20.23  & $-$23 52 35.6 &  $-3.41$ &  6.12  &  $<-1.05$ &  No     &   1, 7                     \\
CS~22888$-$031       &  6553564535381928320   &  23 11 32.48  & $-$35 26 42.9 &  $-3.71$ &  5.64  &  $<-0.71$ &  No     &   1, 15                    \\
HE~2314$-$1554       &  2406838894596713856   &  23 17 01.12  & $-$15 37 49.5 &  $-3.33$ &  5.89  &  $-0.31$  &  No     &   1, 18                    \\
HE~2318$-$1621       &  2406023396270909440   &  23 21 21.50  & $-$16 05 06.0 &  $-3.67$ &  6.30  &  $-1.61$  &  No     &   1, 30                    \\
CS~22949$-$048       &  2632781348624598784   &  23 26 07.41  & $-$05 50 06.7 &  $-3.55$ &  5.81  &  $-1.63$  &  No     &   1, 5                     \\
CS~22949$-$037       &  2634585342263017984   &  23 26 29.80  & $-$02 39 57.9 &  $-4.38$ &  5.97  &  $-0.78$  &  No     &   1, 5                     \\
CS~22957$-$013       &  2446067373532883712   &  23 55 49.05  & $-$05 22 53.0 &  $-2.98$ &  6.17  &  $-0.80$  &  No     &   1, 5                     \\
CS~22957$-$027       &  2447834632315747840   &  23 59 13.14  & $-$03 53 48.4 &  $-3.19$ &  7.87  &  $-1.00$  & Yes     &   1, 5                     \\
\enddata
\tablerefs{(1) \citet{yoon2016}, (2) \citet{placco2016b}, (3) \citet{zepeda2022}, (4) \citet{RAVE6}, (5) \citet{roederer2014}, (6) \citet{hansen2016a}, (7) \citet{arentsen2019}, (8) \citet{cohen2013}, (9) \citet{caffau2025}, (10) \citet{placco2024}, (11) \citet{hollek2011}, (12) \citet{aguado2022}, (13) \citet{gonzalez2020}, (14) \citet{yong2013}, (15) \citet{bonifacio2007}, (16) \citet{limberg2021}, (17) \citet{lai2008}, (18) \citet{barklem2005}, (19) \citet{allen2012}, (20) \citet{latham2002}, (21) \citet{placco2016}, (22) \citet{aoki2013}, (23) \citet{bandyopadhyay2024}, (24) \citet{mardini2019a}, (25) \citet{li2022}, (26) \citet{jacobson2015}, (27) \citet{howes2015}, (28) \citet{RAVE5}, (29) \citet{aoki2018}, (30) \citet{placco2014}}
\tablecomments{
References listed are for stellar parameters, radial velocities, and binarity determinations.}
\end{deluxetable}

\section{CEMP-no statuses}\label{sec:appendix_upper_limits}

\begin{figure*}[ht!]
\centering
\includegraphics[scale=0.53,trim={0 0 3.1cm 0}]{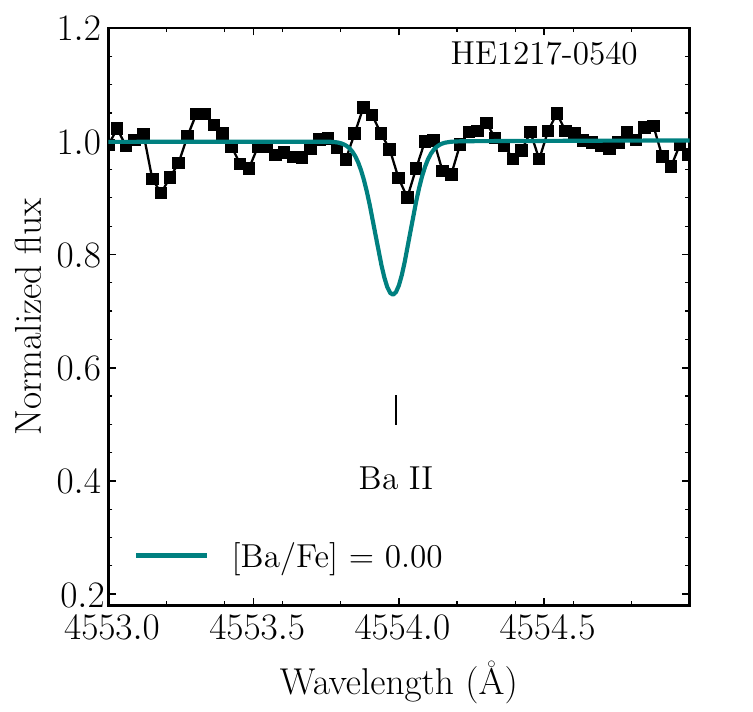}
\includegraphics[scale=0.53,trim={0 0 3.1cm 0}]{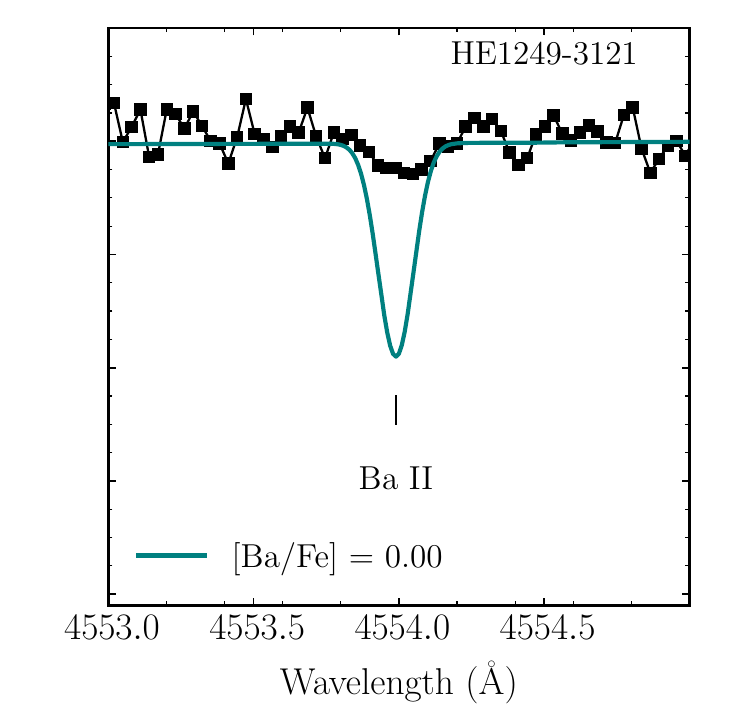}
\includegraphics[scale=0.53,trim={0 0 0 0}]{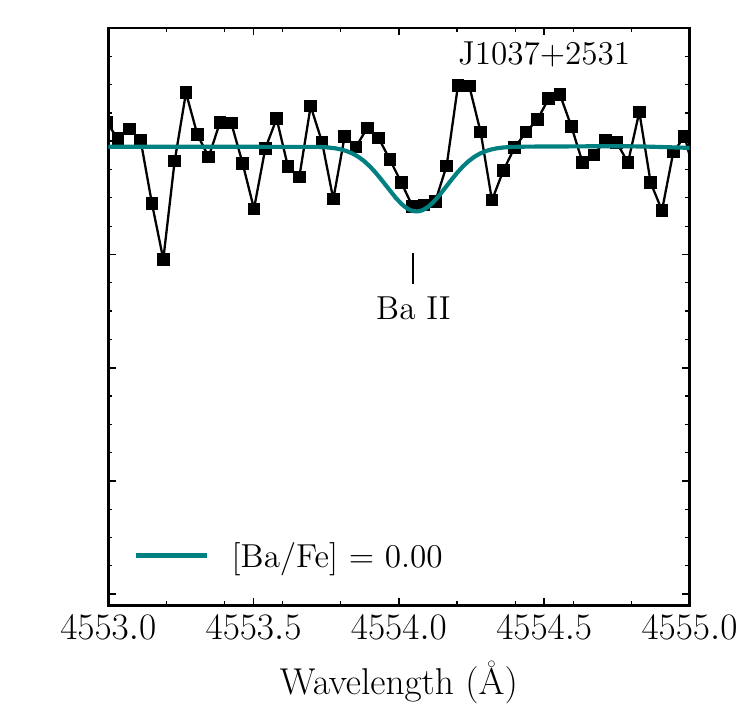}
\caption{\label{fig:ba_synthesis} Archival spectra of HE~1217$-$0540, HE~1249$-$3121, and J1037$+$2531, centered on the \ion{Ba}{ii} line at 4554~\AA. Synthetic spectra with \BaFe\ $=0.0$ are plotted in teal.}
\end{figure*}

For three CEMP-no stars in our sample, no Ba abundance or upper limit is available in the literature. To confirm their CEMP-no statuses, we synthesized the \ion{Ba}{ii} line at 4554~\AA\ using an abundance of \AB{Ba}{Fe} $=0.0$ and compared with existing spectra. Archival spectra of HE~1217$-$0540 and HE~1249$-$3121 were taken from VLT/UVES (program 170.D-0010), and stellar parameters for the synthesis are from \citet{barklem2005}. For HE~1217$-$0540, we use \teff\ $=5700$~K, \logg $=4.20$, \FeH\ $=-2.95$, and \vmic\ $=1.41$~\kms. For HE~1249$-$3121, we use \teff\ $=5373$~K, \logg $=3.40$, \FeH\ $=-3.23$, and \vmic\ $=1.58$~\kms. For J1037$+$2531, we use archival spectra from GEMINI/GRACES (program GN-2019B-Q-219) and stellar parameters \teff\ $=6569$~K, \logg $=4.37$, \FeH\ $=-2.5$, and \vmic\ $=1.3$~\kms\ from \citet{jeong2023}. The synthesis was done in SMHR \citep{casey2025} running the 2017 version of MOOG\footnote{\url{www.github.com/alexji/moog17scat}} \citep{sneden1973,sobeck2011}, using $\alpha$-enhanced (\AB{$\alpha$}{Fe} $=+0.4$) ATLAS9 models \citep{castelli2003}, a Ba line list assembled from \citet{placco2021}, 
and \rproc\ isotopic ratios from \citet{sneden2008}. The results are displayed in Figure \ref{fig:ba_synthesis}, where the black dots are the observed spectra and the teal line is the synthesis. The synthetic spectral lines are stronger than the observed flux, confirming the subsolar \BaFe, and hence the CEMP-no status, of these three stars.

\end{document}